\documentclass[aps,prr,reprint,superscriptaddress,amsmath,amssymb,longbibliography]{revtex4-1}

\usepackage{graphicx}% Include figure files
\usepackage{dcolumn}% Align table columns on decimal point

\usepackage{bm}% bold math
\usepackage[english]{babel}

\usepackage[utf8]{inputenc}
\usepackage[T1]{fontenc}
\usepackage{mathtools}
\usepackage{placeins}

\usepackage{letltxmacro}

\LetLtxMacro{\ORIGselectlanguage}{\selectlanguage}
\makeatletter
\DeclareRobustCommand{\selectlanguage}[1]{%
  \@ifundefined{alias@\string#1}
    {\ORIGselectlanguage{#1}}
    {\begingroup\edef\x{\endgroup
       \noexpand\ORIGselectlanguage{\@nameuse{alias@#1}}}\x}%
}
\newcommand{\definelanguagealias}[2]{%
  \@namedef{alias@#1}{#2}%
}
\makeatother

\definelanguagealias{en}{english}

\newcommand{\rref}[1]{Eq.\ (\ref{#1})}
\newcommand{\rrefsa}[1]{Eqs.\ (\ref{#1})}
\newcommand{\rrefsb}[1]{(\ref{#1})}
\newcommand{\bp}{{\bf p}}
\newcommand{\bq}{{\bf q}}
\newcommand{\bs}{{\bf s}}
\newcommand{\bk}{{\bf k}}
\newcommand{\bkp}{{\bf k'}}
\newcommand{\kint}{k_\mathrm{int}}

\DeclareUnicodeCharacter{2009}{\,}

\usepackage{color}
\usepackage{multirow}

\usepackage[unicode]{hyperref}
\hypersetup{
   unicode=true,          % non-Latin characters in Acrobat??s bookmarks
   plainpages=false,
   colorlinks=true,       % false: boxed links; true: colored links
   citecolor=blue,        % color of links to bibliography
}

\usepackage{url}
\usepackage{todonotes}

\begin{document}

\title{ Eliminating the wave function singularity for ultracold atoms by similarity transformation
}

\date{\today}% It is always \today, today,

\author{P\'eter Jeszenszki}
 %\email{jeszenszki.peter@gmail.com}
 \affiliation{Dodd-Walls Centre for Photonics and Quantum Technology, PO Box 56, Dunedin 9056, New Zealand}
\affiliation{%
New Zealand Institute for Advanced Study, and Centre for Theoretical Chemistry and Physics, Massey University, Private Bag 102904 North Shore, Auckland 0745, New Zealand
}%
\author{Ulrich Ebling}
 \affiliation{Dodd-Walls Centre for Photonics and Quantum Technology, PO Box 56, Dunedin 9056, New Zealand}
\affiliation{%
New Zealand Institute for Advanced Study, and Centre for Theoretical Chemistry and Physics, Massey University, Private Bag 102904 North Shore, Auckland 0745, New Zealand
}%
\author{Hongjun Luo}
%\email{H.Luo@fkf.mpg.de}
\affiliation{
 Max Planck Institute for Solid State Research, Heisenbergstra{\ss}e 1,
70569 Stuttgart, Germany
}%
\author{Ali Alavi}
%\email{A.Alavi@fkf.mpg.de}
\affiliation{
 Max Planck Institute for Solid State Research, Heisenbergstra{\ss}e 1,
70569 Stuttgart, Germany
}%
 \affiliation{Department of Chemistry, University of Cambridge,
Lensfield Road, Cambridge, CB2 1EW, United Kingdom}
\author{Joachim Brand}
% \email{J.Brand@massey.ac.nz}
 \affiliation{Dodd-Walls Centre for Photonics and Quantum Technology, PO Box 56, Dunedin 9056, New Zealand}
\affiliation{%
New Zealand Institute for Advanced Study, and Centre for Theoretical Chemistry and Physics, Massey University, Private Bag 102904 North Shore, Auckland 0745, New Zealand
}%
\affiliation{
 Max Planck Institute for Solid State Research, Heisenbergstra{\ss}e 1,
70569 Stuttgart, Germany
}%

\begin{abstract}
A hyperbolic singularity in the wave-function of $s$-wave interacting atoms is  the root  problem for any accurate numerical simulation. Here we apply the transcorrelated method, whereby the wave-function singularity is explicitly described by a two-body Jastrow factor, and then folded into the Hamiltonian via a similarity transformation.
The resulting non-singular eigenfunctions are approximated by stochastic Fock-space diagonalisation
with energy errors scaling with $1/M$ in the number $M$ of single-particle basis functions.
The performance of the transcorrelated method is demonstrated on the example of strongly correlated fermions with unitary interactions. The current method provides the most accurate ground state energies so far for three and four fermions in a rectangular box with periodic boundary conditions.
\end{abstract}

\maketitle

\section{Introduction}

Quantum gases make the
study of strongly correlated many-body physics accessible \cite{Zwerger2012a}, and can be probed 
with exquisite control in the many-particle \cite{Ku2012,Mukherjee2016,Eigen2018,Carcy2019,mukherjee_spectral_2019} and few-particle \cite{Wenz2013a,Reynolds2020} regimes.
At ultracold temperatures their interactions  are accurately described by  
only the $s$-wave scattering length $a_s$  \cite{weiner_experiments_1999}, or 
no parameters 
in  the universal regime of unitary interactions \cite{Braaten2006}.
Despite this apparent simplicity,
 it is nevertheless a great challenge to represent the complicated many-body wave functions in computational approaches \cite{giorgini_theory_2008}.
Specifically, a $1/r$ divergence when two particles with distance $r$  approach each other  \cite{bethe_h._quantum_1935} introduces divergent short-range correlations into the wave function.
While exact approaches 
are limited to 
four particles
\cite{Busch1998,petrov_weakly_2004,Werner2006,liu_three_2010,deltuva_universality_2017}, computational approaches for larger particle numbers rely on lattice discretisation with renormalised interactions \cite{werner_general_2012} (employed at zero \cite{Carlson2011,endres_lattice_2013,lee_ground_2008,he_superfluid_2019} and finite temperature \cite{lee_cold_2006,burovski_fermihubbard_2006,bulgac_quantum_2008, goulko_thermodynamics_2010, goulko_numerical_2016, rammelmuller_finite-temperature_2018,jensen_contact_2019,jensen_pairing_2020,richie-halford_emergence_2020}), 
the closely related renormalised contact interaction \cite{stetcu_effective_2007},
finite-range pseudopotentials \cite{yin_small_2013,Jeszenszki2018a,Jeszenszki2019}, or the more sophisticated effective Hamiltonian approaches  \cite{Alhassid2008,christensson_effective-interaction_2009}. By introducing an ultraviolet cutoff, these approaches do not accurately describe the short-range correlations, and suffer from slow convergence upon increasing the 
number of lattice sites or basis functions.

In this work we apply the transcorrelated method \cite{boys_determination_1969} to remove the short-range correlations from the wave function by a similarity transformation of the many-body Hamiltonian.
Previously, the transcorrelated method was applied to Coulomb-interacting electrons \cite{hino_application_2002,ten-no_new_2002,luo_combining_2018,cohen_similarity_2019} and to ultracold atoms in one dimension \cite{jeszenszki_accelerating_2018}. In these cases the wave function is non-singular but has a cusp, i.e.\  is continuous with a discontinuous first derivative \cite{kato_eigenfunctions_1957}.
Here, we extend  the transcorrelated approach to the hyperbolic singularity and show that it is completely removed.
The similarity-transformed, transcorrelated Hamiltonian is free from singular zero-range interactions, which are replaced by 
new two-body and three-body terms, and has non-singular eigenfunctions.
The advantages of the method are demonstrated  by ground-state calculations
with stochastic projective diagonalization in Fock-space \cite{booth_fermion_2009}.
For a few fermions with unitary interactions 
we find that the error of the energy is the smallest among the available methodologies. Moreover, this error decays with $1/M$, where $M$ is the number of the single-particle plane wave basis functions.
This is the fastest convergence rate so far.

This paper is organized as follows: Section \ref{sec:theory} introduces the correlation factor for zero-range $s$-wave interactions and the transcorrelated transformation of the many-body Hamiltonian. Results on three and four fermions are described in Sec.~\ref{sec:results} before concluding the main text with Sec.~\ref{sec:conclusions}. Appendices provide many derivations and details starting with the real-space form of the correlation factor in App.~\ref{sec:appendixa}. Appendix \ref{sec:trcorrFH} provides a derivation to show that the matrix elements of the similarity transformed Fermi-Huang pseudopotential between smooth functions vanish. The smoothness of the transcorrelated two-particle eigenfunctions is examined in App.~\ref{sec:appendixC}. Appendix \ref{sec:appendixD} concerns the transcorrelated many-body Hamiltonian and provides it in second-quantized form, while App.~\ref{sec:appendixE} describes the algorithm used for evaluating an infinite sum that appears in the transcorrelated Hamiltonian. Appendix \ref{sec:appendixF} finally provides details of the numerical calculations including parameters relating the the FCIQMC method (App.~\ref{sec:fciqmc}), extrapolation procedures (App.~\ref{sec:extrapol})  and data (App.~\ref{sec:4val}), details about lattice renormalization procedures used for comparison (App.~\ref{sec:disp}), and about the exact and and approximate implementation of three-body interaction terms (App.~\ref{sec:threebodyterms}).

\section{Correlation factor and similarity transformation} \label{sec:theory}

Zero-range $s$-wave interactions
are characterized
by the Bethe-Peierls boundary condition \cite{bethe_h._quantum_1935}
\begin{align} \label{eq:BethePeierls}
\Psi({\bf r}_1, {\bf r}_2, \dots) \sim \frac{1}{r_{ij}} - \frac{1}{a_s} + \mathcal{O}(r_{ij}) \quad \textrm{for} \quad r_{ij}\to 0,
\end{align}
where  $r_{ij}=|{\bf r}_{i}-\bf{r}_j|$ is the distance between 
particles $i$ and $j$, and $\Psi$ is the  many-body wave function.
We aim to deal with the divergent short-range part with a Jastrow factor $e^\tau$ by writing
\begin{eqnarray}
\Psi({\bf r}_1, {\bf r}_2, \dots) &=& e^{\tau({\bf r}_1, {\bf r}_2, \dots)} \Phi({\bf r}_1, {\bf r}_2, \dots) , \label{trcorransatz}
\end{eqnarray}
which defines the transcorrelated wave function $\Phi$.
$\tau$ is chosen as a sum of pair correlation factors 
$\tau({\bf r}_1, {\bf r}_2, \dots) = \sum_{i<j} u \left( r_{ij} \right)$.
Requiring 
\begin{align}
    u(r) \sim \mathrm{const.} -\ln \left(\frac{r}{a_s}\right) -\frac{r}{a_s}+\mathcal{O}\left( r^2\right) , \label{ushortrange}
\end{align}
allows the Jastrow factor to carry the main part of the singular short-range correlation and
leaves the transcorrelated wave function $\Phi$ non-singular.
Inserting the ansatz \eqref{trcorransatz} into the Schr\"o\-din\-ger equation $H\Psi = E \Psi$ and multiplying it with $e^{-\tau}$ from the left yields 
\begin{align} \label{eq:trcorr}
\tilde{H} \Phi = E \Phi ,
\end{align}
where $\tilde{H} = e^{-\tau} H e^\tau$ is the transcorrelated Hamiltonian \cite{boys_determination_1969}. 

It is convenient to define the correlation factor in momentum space, and we choose
\begin{align}
\tilde{u}(k) &=
 \begin{cases}
\frac{2 \pi^2}{k^3} + \frac{8 \pi}{a_s k^4}    & \text{if } k \ge k_c \ , \\
0             & \text{if } k < k_c \ ,
\end{cases} \label{corrfactorop4}
\end{align}
where $k_c$ is a  momentum cutoff. The real-space correlation factor is obtained by Fourier transform $u(r) = (2\pi)^{-2}\int_0^\infty dk\,\tilde{u}(k) k \sin(kr)/r$ and the corresponding Jastrow factor $\exp(u)$ is shown in Fig.\ \ref{fig:Jastrow}. More details are provided  in Appendix \ref{sec:appendixa},
where it is shown that $u(r)$ satisfies 
Eq.\ (\ref{ushortrange}).
The momentum cutoff $k_c$ 
damps out the real-space
$u(r)$ for large $r$. The idea 
is that long-range correlations in the transcorrelated wave function $\Phi$ can be effectively dealt with by 
the expansion in a Fock basis, as we will show, while the Jastrow factor $e^\tau$ very efficiently removes the singular short-range correlations.

 \begin{figure}
\begin{center}
\includegraphics[scale=0.7]{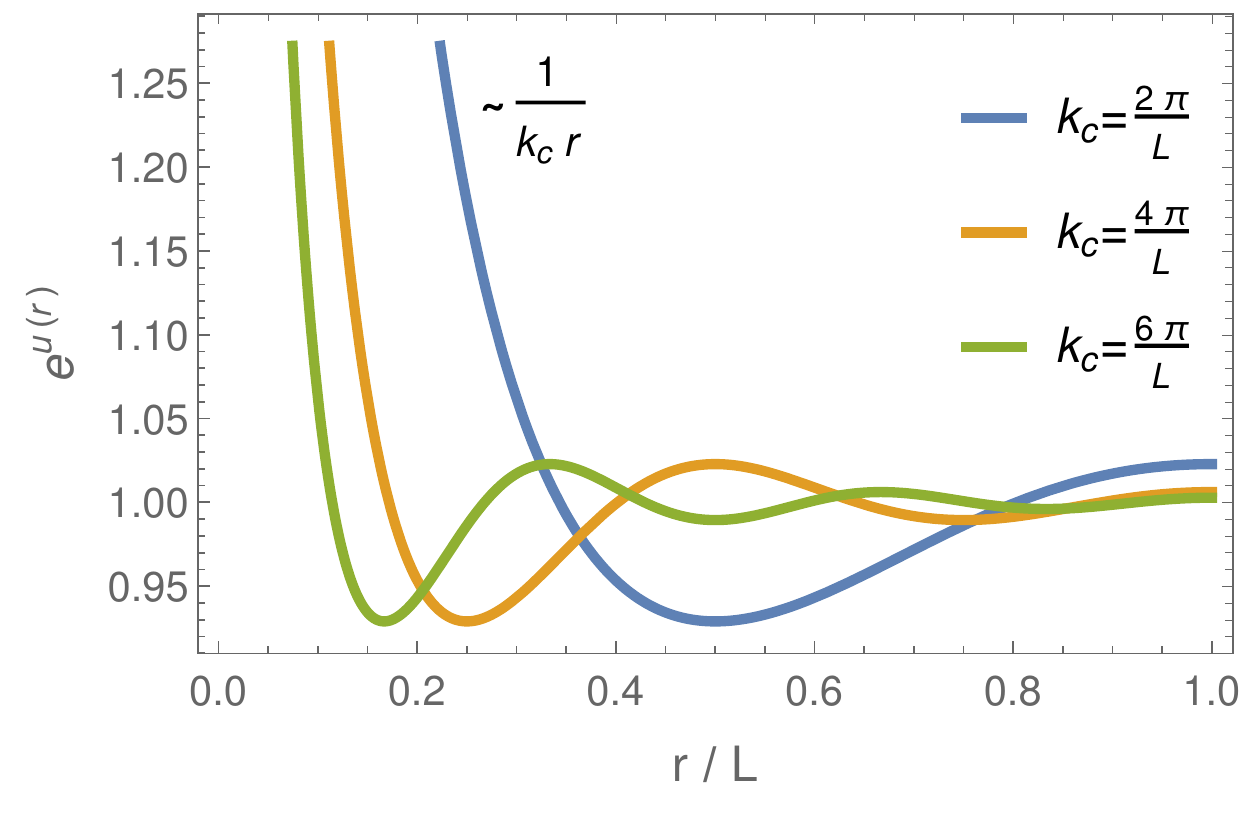}
\caption{The Jastrow factor $e^{u(r)}$ as a function of particle separation $r$ with unitary interaction at different values of the momentum cutoff. $L$ is the size of the computational box. \label{fig:Jastrow}}
\end{center}
\end{figure}

For definiteness we consider a system of ultracold atoms of mass $m$ 
with Hamiltonian $H = H_1 + V_{\mathrm{FH}}$,
where $H_1 = \sum_i -\frac{\hbar^2}{2m} \nabla_i^2 + V_\mathrm{trap}$ is the single-particle part with trapping potential $V_\mathrm{trap}$.
The
zero-range $s$-wave interactions between atoms are represented by the Fermi-Huang pseudopotential \cite{huang_quantum-mechanical_1957}
\begin{eqnarray}
V_{\mathrm{FH}} &=& g \sum_{i<j} \delta \left( {\bf r}_{ij} \right)
\frac{\partial}{\partial r_{ij}} r_{ij} \ , \label{defVFH}
\end{eqnarray}
where  $g = 4\pi\hbar^2 a_s/m$ is the potential strength. The derivative term regularizes the otherwise pathological contact interaction and enforces the Bethe-Peierls boundary conditions of Eq.\ \eqref{eq:BethePeierls}  \cite{huang_quantum-mechanical_1957,yi_physics_2018}.
This pseudopotential has been applied in exact \cite{Busch1998,Olshanii2001} and perturbative \cite{huang_quantum-mechanical_1957} treatments, but it has a limitation in the Fock-state based approaches. As the Fock-state basis functions are smooth, the Fermi-Huang pseudopotential reduces to a simple Dirac-delta function, which is pathological in two and three dimensions \cite{esry_validity_1999, rontani_renormalization_2017,doganov_two_2013,stetcu_effective_2007}. It is suitable for use with the transcorrelated method, however, as long as the Jastrow factor $e^\tau$ is designed to fulfil
Eq.\ \eqref{eq:BethePeierls}.

The similarity transformation $\tilde{H} = e^{-\tau} H e^\tau$ is applied term by term and does not change simple functions of the coordinates because the correlation factor is local in the coordinates.
The kinetic energy and $V_{\mathrm{FH}}$
contain coordinate derivatives and thus generate additional terms.
Specifically, $e^{-\tau} V_{\mathrm{FH}}  e^{\tau} =  V_{\mathrm{FH}} + [V_{\mathrm{FH}} , \tau]$ and,  as we show in Appendix \ref{sec:trcorrFH},
\begin{align}
 \langle \chi|  [V_{\mathrm{FH}} , \tau]  |\phi \rangle =& \langle \chi| g \sum_{i<j} \delta \left( {\bf r}_{ij} \right)
\frac{\partial u(r_{ij})}{\partial r_{ij}} r_{ij}  |\phi \rangle , \label{eq:tcFH}
\end{align}
for  wave functions $\phi$ and $\chi$ that are bounded and have bounded first derivatives.
In Appendix \ref{sec:trcorrFH} it is shown that the matrix elements of the
similarity transformed Fermi-Huang pseudopotential $ \langle \chi|   e^{-\tau} V_{\mathrm{FH}}  e^{\tau}|\phi \rangle$
vanish due to  cancellation 
as long as the correlation factor $u(r)$ is chosen to have the appropriate short-range asymptotics of
Eq.\ \eqref{ushortrange}.
Thus,
the singular pseudopotential is removed
and the transcorrelated Schr\"odinger equation \eqref{eq:trcorr} can be solved with a non-singular wave function $\Phi$.
This insight presents the
main result of this Letter.

The transcorrelated Hamiltonian still acquires terms that originate from the kinetic energy operator, and finally reads
\begin{align}
\tilde{H} = H_1 - \sum_i \left[ \frac{1}{2} \nabla_i^2 \tau  + \left( \nabla_i \tau \right) \nabla_i + \frac{1}{2} \left( \nabla_i \tau \right)^2 \right] \frac{\hbar^2}{m} .\label{trcorrHam}
\end{align}
The new terms 
represent an effective interaction potential that is less singular than the Fermi-Huang pseudopotential.
The leading singular term is $\left( \nabla_i \tau \right) \nabla_i \sim \left( \sum_j {\bf r}_{ij} / r_{ij}^2 \right) \nabla_i$, which  has a $1/r$ divergence and is also non-hermitian. Similar to the Coulomb potential it leads to a cusp feature, where the transcorrelated wave function $\Phi$ is continuous with a discontinuous first derivate (see in Appendix \ref{sec:appendixC}). 
In momentum space, $\Phi$ thus decays with $1/k^4$ for large $k$ instead of $1/k^2$ for the original wave function $\Psi$. It is this rapid decay for large $k$ that makes it feasible to expand the problem in a plane-wave basis without the need for a renormalized (or running) coupling constant. The second-quantized form of the transcorrelated Hamiltonian in momentum space is presented in  Appendix \ref{sec:appendixD}.

Although the similarity transformation eliminates the singularity from the wave function without modifying the spectrum of the Hamiltonian, it introduces new challenges for numerical calculations. The non-hermitian term $\left( \nabla_i \tau \right) \nabla_i$ prevents in general the variational minimization of the energy. However, the ground-state energy can still be found by projection techniques as proved in Ref.~\cite{luo_combining_2018} and previously demonstrated in Refs.~\cite{luo_combining_2018,cohen_similarity_2019,jeszenszki_accelerating_2018,dobrautz_compact_2019}.
As a consequence of the non-hermiticity, the approximate energies no longer provide an upper bound to the exact ground state energy.
Non-hermitian terms are common in the transcorrelated and coupled-cluster methods and usually do not cause problems.
Hypothetically, the projection onto a finite basis could
lead to pairs of complex conjugate eigenvalues with small imaginary parts in the vicinity of a accidental eigenvalue degeneracies. The stochastic projection used in this work would then fail to fully converge and resolve the near-degenerate eigenvalues. This is easy to diagnose but we have not encountered this situation so far. A removal strategy for complex eigenvalues was suggested in the context of coupled cluster theory 
\cite{Kohn2007}. 

The terms $ \frac{1}{2} \nabla_i^2 \tau$ and $\frac{1}{2} \left( \nabla_i \tau \right)^2$ 
have leading $1/r_{ij}^2$ and $1/r_{ij}r_{ik}$ contributions, respectively, and partly compensate each other but leave an uncompensated three-body attraction. This long-range interaction represents a mediated three-body attraction that is familiar from Efimov physics \cite{Efimov1970,Naidon2017}. It  permits three-body bound states for resonantly interacting bosons but not for fermions.
Three-body interactions are common in the transcorrelated method and 
were previously implemented for the Hubbard model \cite{dobrautz_compact_2019} and 
electronic structure calculations in atoms \cite{cohen_similarity_2019}. 
Appendix \ref{sec:threebodyterms} presents details of the efficient numerical implementation of the three-body terms as well as a  two-body approximation that saves up to a factor four in computer time while still maintaining high accuracy.

\section{Numerical results}\label{sec:results}

For numerical calculations the transcorrelated Hamiltonian $\tilde{H}$ is expanded as a finite matrix in a  Fock basis of antisymmetrized products of single-particle plane waves with a momentum cutoff.
For two particles the ground state energy is calculated with (numerically) exact diagonalization. For three and four fermions the full matrix diagonalization was not possible due to the enormous size of Hilbert space. Hence we used a stochastic projection method known as  Full Configuration Interaction Quantum Monte Carlo (FCIQMC) \cite{booth_fermion_2009,booth_linear-scaling_2014} to obtain the ground state energies. A  description of 
the method is provided in Appendix \ref{sec:appendixF}.

\begin{figure}
\begin{center}
\includegraphics[scale=0.7]{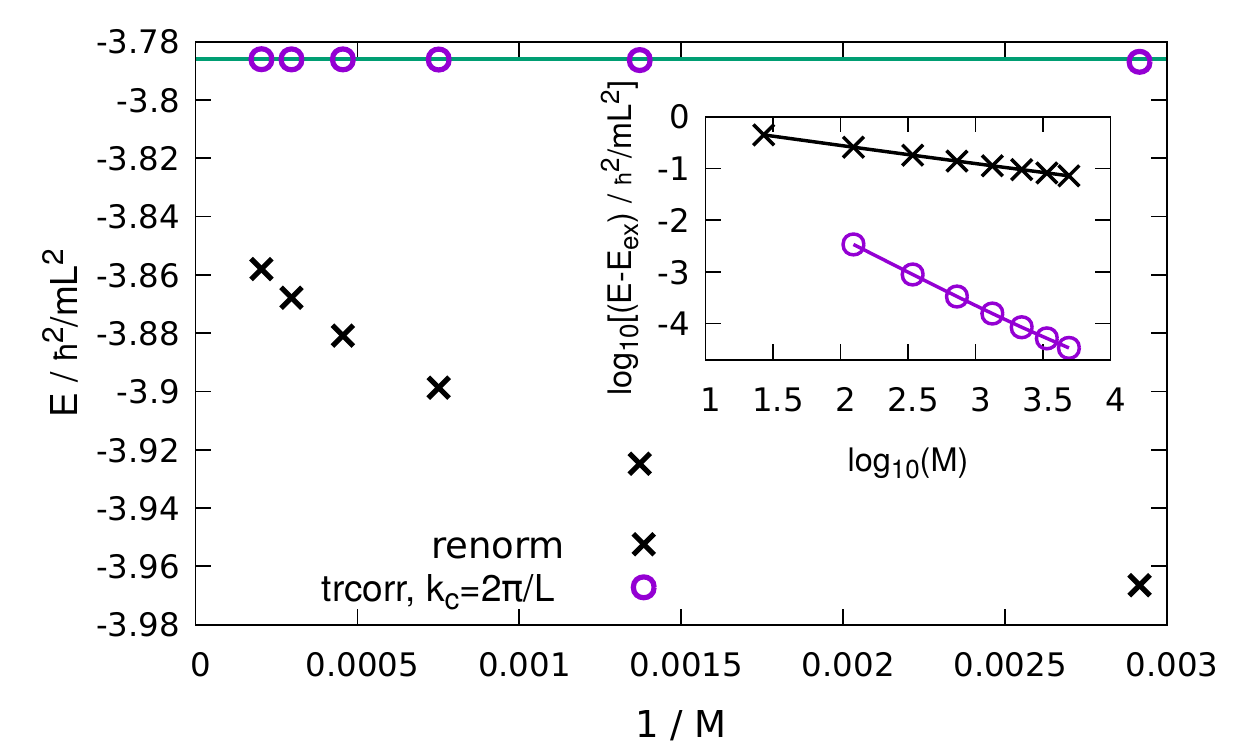}
\caption{Ground-state energy of two particles with unitary interactions ($1/a_s = 0$) vs.\  inverse size of single-particle basis $1/M$ from the transcorrelated method (circles) and with renormalized Dirac delta (crosses). Green solid line: Reference energy $E_\mathrm{ex}=-3.786005 \, \hbar^2/mL^2$ 
\cite{pricoupenko_three_2007}. 
Inset: 
Difference to the reference energy on a log-log plot
indicating power-law scaling 
$\propto 1/M^{1/3}$ for  renormalization  and  $\propto 1/M$ for the transcorrelated approach.
\label{fig:absenerg2part}}
\end{center}
\end{figure}

Results for two particles 
are presented in  Fig.\ \ref{fig:absenerg2part}.
The energy is shown as a function of the inverse of the basis set size $M$, where $\sqrt[3]{M}$ is the number of single-particle plane-wave basis functions per linear dimension of the cube. Hence, the zero on the $x$-axis represents the complete basis set limit.
The transcorrelated energies are compared to standard lattice renormalization \cite{werner_general_2012}, where a running coupling constant $g_0$ is scaled with the number of lattice points $M$ as
$g_0^{-1} = m/4\pi\hbar^2 a_s - m K M^{1/3}/4\pi \hbar^2L$
 ($K=2.442749607806335...$) \cite{Castin2006}. It is not only seen that the transcorrelated method gives smaller errors by orders of magnitudes for the same $M$, but also that scaling of the errors with $M$ follows a faster power law decay.
For the renormalization approach, we find a scaling of $M^{-1/3}$ consistent with the previous results from lattice calculations \cite{werner_general_2012,Carlson2011,bour_precision_2011}.
In the transcorrelated approach the error decays with $M^{-1}$. This is the same scaling as obtained for Coulomb-interacting systems, e.g.\ the homogeneous electron gas, which is consistent with the Coulomb-like nature of the transcorrelated Hamiltonian. The $M^{-1}$ scaling is the fastest scaling we found in the literature, and is shared, e.g.\ with the improved lattice action used for Auxiliary Field Quantum Monte Carlo (AFQMC) calculations by Endres \emph{et al.} \cite{endres_lattice_2013} or the renormalized lattice Hamiltonian with ``magic'' dispersion relation discussed in Ref.\ \cite{werner_general_2012}.

\begin{figure}
\begin{center}
\includegraphics[scale=0.7]{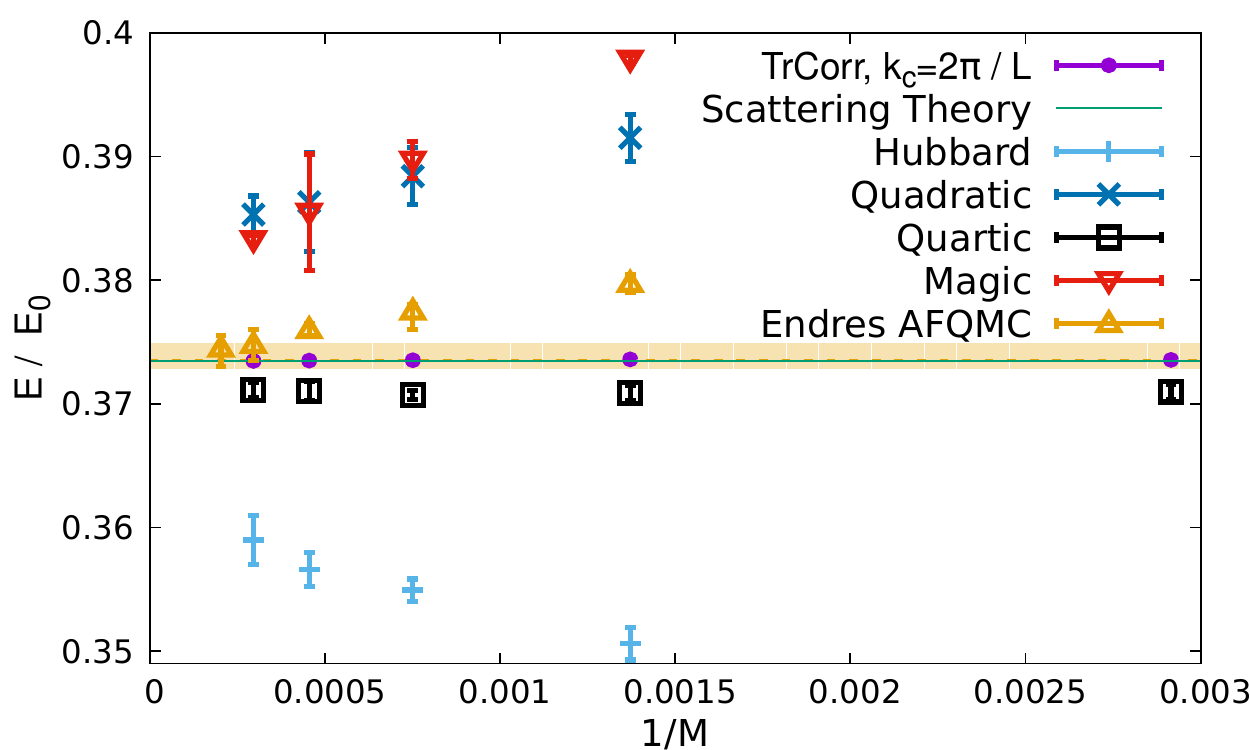}
\caption{Energy of zero-momentum ground state of two spin-up and one spin-down fermions with unitary interactions vs.\  inverse 
size of single-particle basis $1/M$.
Transcorrelated  (``TrCorr'') are compared with semi-analytical results from Ref.\ \cite{Note2} (``Scattering Theory'') and AFQMC 
(``Endres AFQMC'')  \cite{endres_lattice_2013}. The yellow band marks the standard error of the extrapolated AFQMC results. 
Renormalised lattice calculations with FCIQMC
using different single-particle dispersions: ``Hubbard'', ``Quadratic'' and ``Quartic'' as in Ref.\ \cite{Carlson2011}
and the ``Magic'' dispersion from Eqs.\ (122) and (124) in Ref.\ \cite{werner_general_2012}. $E_0=4\pi^2\hbar^2/mL^2$ is the non-interacting energy.
\label{fig:absenerg3part}}
\end{center}
\end{figure}

Results for three fermions in the lowest energy state with zero total momentum\ \footnote{The true ground state has finite momentum \cite{yin_small_2013}.} are 
shown in Fig.\ \ref{fig:absenerg3part}. While the $M^{-1}$ scaling 
can be observed for the ``Endres AFQMC'' values, 
the transcorrelated results are much more accurate already  for very modest basis set size and hardly distinguishable from the reference values on the scale of the figure.
Moreover, in Appendix \ref{sec:threebodyterms} we show that approximate calculations avoiding the numerically expensive three-body excitations achieve the same accuracy within our statistical errors. 

Figure  \ref{fig:absenerg3part} also shows renormalized lattice calculations with different single-particle dispersion relations as discussed in Ref.\   \cite{werner_general_2012,Carlson2011} obtained with FCIQMC. Since they are expected to show slower scaling than $M^{-1}$, the energy dependence does not appear linear in Fig.\  \ref{fig:absenerg3part}. The renormalized lattice method scales with $M^{-1/3}$ when using a Hubbard, quadratic or quartic dispersion, and $M^{-2/3}$ for a ``magic'' dispersion \cite{werner_general_2012}.

\begin{figure}
\begin{center}
\includegraphics[scale=0.7]{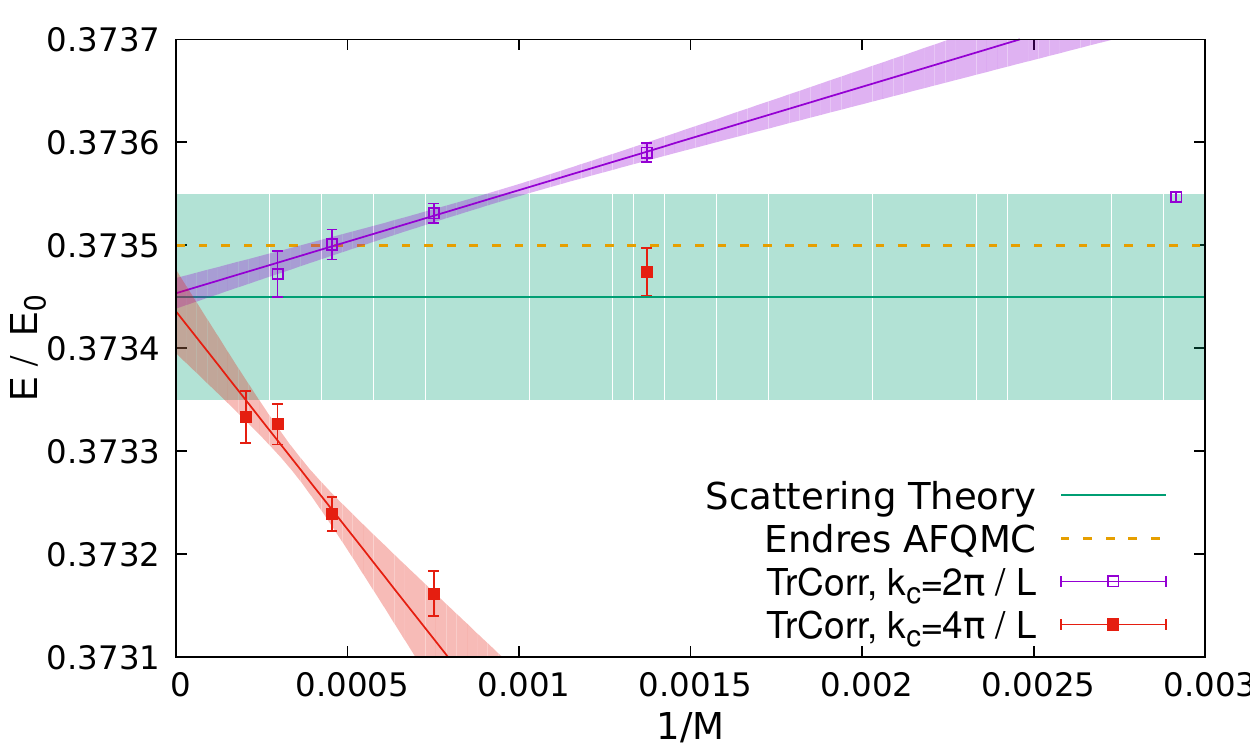}
\caption{
Detail from Fig.\ \ref{fig:absenerg3part} at enlarged scale. The error bars of the transcorrelated results show the stochastic errors from the FCIQMC method. The reference value from Ref.\ \cite{Note2} (``Scattering Theory'') is shown with error band in green,  extrapolated result from ``Endres AFQMC''  \cite{endres_lattice_2013} as dashed (yellow) line (error not shown).
The purple and red bands indicate the $1\sigma$ confidence bands  obtained from $\chi^2$ fitting of the transcorrelated FCIQMC (``TrCorr") results (four largest $M$ values). For details  see  Appendix \ref{sec:extrapol}.
\label{fig:3b}}
\end{center}
\end{figure}

The transcorrelated energies for three particles are shown again in Fig.\  \ref{fig:3b}  with a magnified energy scale and with different momentum cutoffs $k_c$ in the correlation factor of Eq.\ \eqref{corrfactorop4}. It is seen that the asymptotic regime of $M^{-1}$ scaling of the energy error is only reached for the larger basis set sizes.
With the known asymptotic scaling properties we can determine the  energies in the infinite basis set limit by extrapolation. 
The extrapolations with two different $k_c$ values are seen to be consistent with each other as well as with the literature results from scattering theory \footnote{The accurate three-particle energy with standard error was received by private communication from Ludovic Pricoupenko and Yvan Castin based on  Ref.~\cite{pricoupenko_three_2007}.} and AFQMC \cite{endres_lattice_2013}, while they have much smaller error bars than previous results. As the final value for the lowest energy with zero total momentum for three fermions we obtain $E/E_0=0.373453 \pm 0.000034$ using $k_c=2 \pi/L$, where $E_0$ is the three-particle energy without interaction between the particles. Compared to the results of Endres  \emph{et al.} \cite{endres_lattice_2013} of  $E/E_0=0.3735 (+0.0014
/-0.0007)$ the error is reduced by more than an order of magnitude.

\begin{figure}
\begin{center}
\includegraphics[width=\columnwidth]{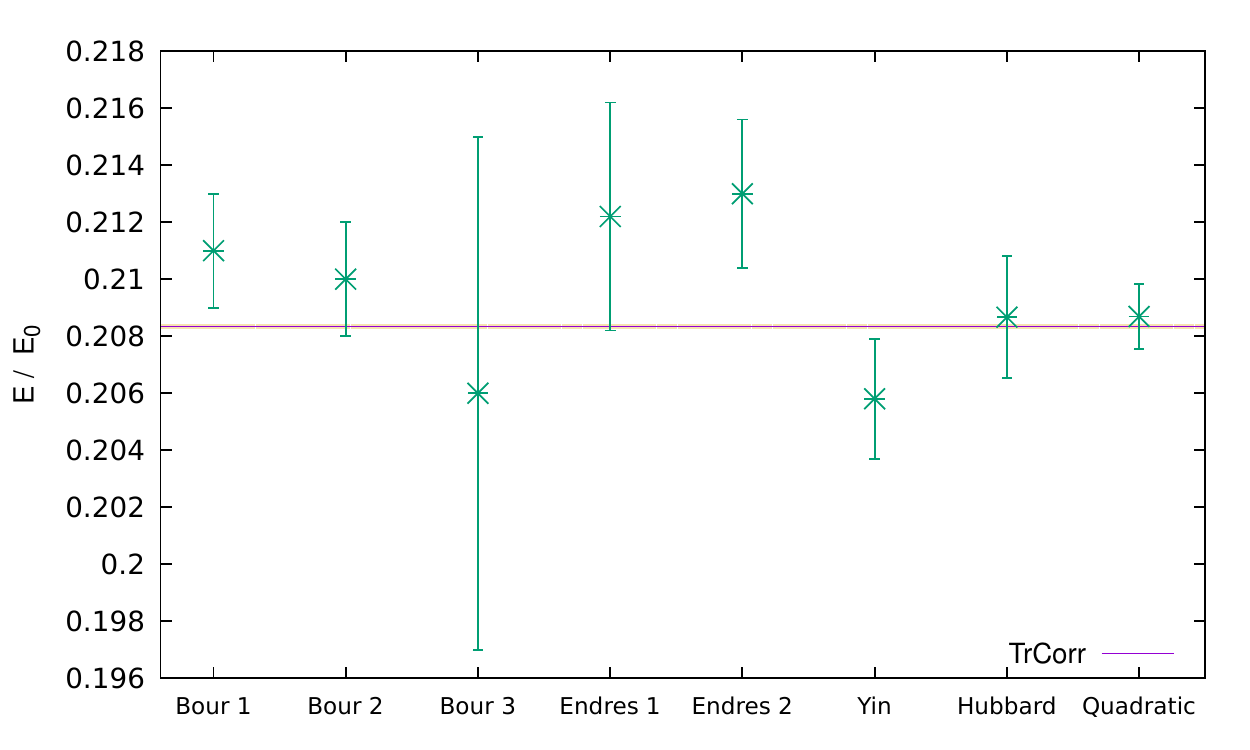}
\caption{Ground-state energy of four fermions extrapolated to  infinite basis set limit.
The horizontal (purple) line (``TrCorr'')  shows the  transcorrelated result $E/E_0 = 0.208339\pm0.000094$
with the error indicated by yellow band.
Results from Ref.\
\cite{bour_precision_2011} with Hamiltonian lattice 1 (``Bour 1'') and 2 (``Bour 2'') and 
AFQMC with Euclidian lattice (``Bour 3'') are shown alongside 
AFQMC results from Ref.\
\cite{endres_lattice_2013} with $\mathcal{O}(4)$ (``Endres 1'') and $\mathcal{O}(5)$ (``Endres 2'') scaling, explicitly correlated Gaussian  (``Yin'') \cite{yin_small_2013}, and
renormalized lattice calculations following Ref.\ \cite{Carlson2011} using ``Hubbard'' and ``Quadratic'' dispersion relations. For the numerical values see Table \ref{tab:fourparticle} in  Appendix \ref{sec:appendixF}.
\label{fig:4b}}
\end{center}
\end{figure}

The results from transcorrelated and renormalized calculations for a four-fermion system are shown in Fig.\ \ref{fig:4b}, where they are compared to literature results
with lattice discretization (exact diagonalization and AFQMC) and explicitly correlated basis set approaches.
Despite the several orders-of-magnitude larger Hilbert-space ($\sim 10^{14}$) we obtain bias-free results from FCIQMC by using the initiator approximation and bias removal by increasing the walker number \cite{cleland_communications:_2010}. The combined stochastic and extrapolation error of the transcorrelated approach is about two orders of magnitude smaller than the best existing literature values (for details and numerical values see Appendix \ref{sec:appendixF}). This result showcases the significance of an explicit treatment of the wave function singularity for improving the accuracy of numerical calculations.

\section{Conclusions} \label{sec:conclusions}

The approach presented here can be easily extended to include trapping potentials or external gauge fields.
The transcorrelated method is thus well suited for highly precise calculations on correlated few-atom systems in microtraps  \cite{Wenz2013a,Reynolds2020}. Extensions to larger particle numbers are feasible and have already been demonstrated with FCIQMC in weakly-correlated regimes \cite{Booth2013}, while recent developments of FCIQMC like the adaptive shift method \cite{Ghanem2019} can help in strongly-correlated regimes. 
The approach can be applied to low dimensional systems, and while it has already proven useful in one dimension  \cite{jeszenszki_accelerating_2018}, the two-dimensional case is an objective for the future. The transcorrelated method could also be employed  with extensions of FCIQMC for finite-temperature calculations with density-matrix Monte Carlo \cite{blunt_density-matrix_2014}, or real time evolution of closed \cite{Guther2017a} or open quantum systems \cite{Nagy2018}.
Beyond the specific numerical approach, we expect that the transcorrelated formalism brings new insight into the treatment of the singularity in the wave function and that it provides a useful theoretical tool in other perturbative and exact computational approaches.

\begin{acknowledgments}
The authors thank Elke Pahl, Andrew Punnett, and Pablo López Riós for discussions, Ludovic Pricoupenko and Yvan Castin for raw data relating to Ref.\ \cite{Note2}, Doerte Blume for a careful reading of the manuscript and for pointing out Ref.\ \cite{yin_small_2013},  and Kai Guther for the help with the integration of code into the NECI program package. PJ thanks the Max Planck Institute for Solid State Research for hospitality during a research stay where this project was started.
This work
was supported by the Marsden Fund of New Zealand (Contract No.\ MAU1604), from government funding managed by
the Royal Society Te Apārangi. We also acknowledge support by the NeSI high-performance computing facilities.
\end{acknowledgments}

\appendix

\section{The correlation factor in real space \label{sec:appendixa}}

In this section, we examine the real-space form of the correlation factor,
\begin{align}
\tilde{u}(k) &=
 \begin{cases}
\frac{2 \pi^2}{k^3} + \frac{8 \pi}{a_s k^4}    & \text{if } k \ge k_c \ , \\
0             & \text{if } k < k_c \ .
\end{cases} \label{corrfactapprox}
\end{align}
The Fourier-transform of function \rrefsb{corrfactapprox} can be calculated analytically,
\begin{align*}
    u(r)&=\int\limits_{-\infty}^\infty  \mbox{d}k^3\; \tilde{u}(|\mathbf{k}|) \frac{e^{i \mathbf{k}\cdot\mathbf{r}}}{(2\pi)^3}=v(r)+\frac{8\pi}{a_s}w(r)  \ ,\\
    v(r)&=\frac{\sin(k_c r)}{k_c r} - \mathrm{Ci}(k_c r) \ , \\
    w(r)&=\frac{\cos(k_c r)}{4 k_c \pi^2} +\frac{\sin(k_c r)}{4 k_c^2 \pi^2 r} + \frac{r}{4 \pi^2}\mathrm{si}(k_c r) \ ,
\end{align*}
Where $\mathrm{Ci}(x) = -\int_x^\infty \frac{\cos(t)\, dt}{t}$ is the cosine integral and $\mathrm{si}(x) = -\int_x^\infty \frac{\sin(t)\, dt}{t}$ is the sine integral.
The boundary condition can be reproduced by expanding  function $v(r)$ and $w(r)$ in Taylor series around $r=0$,
\begin{align*}
    v(r) &= - \ln \left( k_c r \right) + 1- \gamma + \mathcal{O}\left( k_c^2 r^2 \right) \ , \\
    w(r) &= \frac{1}{2 \pi^2 k_c} - \frac{r}{8 \pi} + \mathcal{O}\left( k_c^2 r^2 \right) \ , \\
    u(r) &= - \ln \left( k_c r \right) + 1- \gamma +  \frac{4}{\pi a_s k_c} -  \frac{r}{a_s} + \mathcal{O}\left( k_c^2 r^2 \right) \ ,
\end{align*}
where $\gamma$ is the Euler-Mascheroni constant. Calculating $e^{u(r)}$, we obtain back the hyperbolic singularity for the Jastrow-factor,
\begin{align*}
e^{u(r)} &= {e^{1-\gamma+\frac{4}{a_s k_c \pi} }}
\left[ \frac{1}{k_c r} - \frac{1}{k_c a_s} + \mathcal{O}(k_c r)   \right] \ .
\end{align*}

\section{Matrix elements of the transcorrelated Fermi-Huang pseudopotential \label{sec:trcorrFH}}

We consider the matrix element of the transcorrelated Fermi-Huang pseudopotential and show that it vanishes, if evaluated with wave functions that are bounded and  have a bounded first derivative 
almost everywhere.

In order to show that, let us consider the transcorrelated Fermi-Huang pseudopotential, 
\begin{align} \label{eq:trFH}
   e^{-\tau} V_{\mathrm{FH}} e^{\tau} &=   V_{\mathrm{FH}} +\left[  V_{\mathrm{FH}} , \tau \right] \ ,
\end{align}
where the commutator can be evaluated if we apply the substitution $V_{\mathrm{FH}}=g\sum_{i<j}  \delta \left( {\bf r}_{ij} \right) \frac{\partial}{\partial r_{ij}} r_{ij}$, 
\begin{align}
\left[  V_{\mathrm{FH}} , \tau \right] = g\sum_{i<j}  \delta \left( {\bf r}_{ij} \right) \frac{\partial \tau }{\partial r_{ij}} r_{ij} \ . \label{FHcommut}
\end{align}
The partial derivative with respect to the separation $r_{ij}=|{\bf r}_{i}-\bf{r}_j|$ is defined in the usual way, where both particles $i$ and $j$ move while the center-of-mass $\frac{1}{2}({\bf r}_{i}+\bf{r}_j)$ is held constant, as are the orientation of the vector $\mathbf{r}_{ij} ={\bf r}_{i}-\bf{r}_j$, and all other particle coordinate vectors $\mathbf{r}_k$ for $k\ne i, k\ne j$. Since $\tau$ depends on the separations of all particle pairs, the chain rule will generate many term, most of which, however, vanish.

In order to evaluate the derivative $\frac{\partial \tau }{\partial r_{ij}}$,
let us substitute in the expansion of $\tau$ in pair correlation functions,
\begin{align}
    \tau = \sum_{i<j} u \left( r_{ij} \right) \ , \label{tauexpofu} 
\end{align} 
into $\frac{\partial \tau }{\partial r_{ij}}$,
\begin{eqnarray}
 \frac{\partial \tau }{\partial r_{ij}}  &=&  \frac{\partial u(r_{ij}) }{\partial r_{ij}}+ \sum_l^{i<l \ne j} \frac{\partial u(r_{il}) }{\partial r_{ij}} + \sum_k^{i \ne k<j } \frac{\partial u(r_{kj}) }{\partial r_{ij}} + \label{tauderiv} \\
 && \hspace{4.5cm} + \sum_{k<l}^{i \ne k, j \ne l} \frac{\partial u(r_{kl}) }{\partial r_{ij}} \ . \nonumber
\end{eqnarray}
The last term on the right-hand side is zero as $r_{kl}$ does not depend on $r_{ij}$. For the second and the third terms on the right-hand side we can apply the chain rule,
\begin{eqnarray}
\frac{\partial u(r_{il})}{\partial r_{ij}} &=& 
 \frac{\partial u(r_{il})}{\partial r_{il}} \underbrace{\sum_{p=1}^3 \frac{\partial r_{il}}{\partial r_{i,p}} \frac{\partial r_{i,p}}{\partial r_{ij}}}_{\cos \theta_{jil}/2} \ , \label{parder1} \\
\frac{\partial u(r_{kj})}{\partial r_{ij}} &=& 
\frac{\partial u(r_{kj})}{\partial r_{kj}} \underbrace{\sum_{p=1}^3 \frac{\partial r_{kj}}{\partial r_{j,p}} \frac{\partial r_{j,p}}{\partial r_{ij}}}_{\cos{\theta_{ijk}}/2} \ , \label{parder2}
\end{eqnarray}
where index $p$ goes through the three spatial directions and $\theta_{jil}$ is the angle between ${\bf r}_{ji}$ and ${\bf r}_{il}$. 

Using the short-range behavior of the correlation factor,
\begin{eqnarray}
u(r) = 1 -\ln \left(\frac{r}{a_s} \right) -\frac{r}{a_s}+\mathcal{O}\left( r^2\right) \ ,
\label{ushortrangeapp}
\end{eqnarray}
the first derivative of $u(r)$ can be evaluated for short interparticle separations,
\begin{eqnarray}
\frac{\mbox{d} u (r)}{\mbox{d} r} &=& -\frac{1}{r}-\frac{1}{a_s}+\mathcal{O}\left(r\right) \ . \label{uderiv}
\end{eqnarray}
Substituting \rrefsa{parder1}-\rrefsb{uderiv} into  \rref{tauderiv}, the explicit expression can be obtained for $\partial \tau / \partial r_{ij}$,
\begin{eqnarray}
\frac{\partial \tau}{\partial r_{ij}} &=& -\frac{1}{r_{ij}}-\sum_l^{i<l \ne j} \frac{\cos \theta_{jil}}{2 r_{il}}  -\sum_k^{i \ne k < j} \frac{\cos \theta_{ijk}}{2 r_{kj}} - \label{taufinalform} \\
&& - \frac{1}{a_s} \left( 1+\sum_l^{i<l \ne j} \frac{\cos \theta_{jil}}{2}  +\sum_k^{i \ne k < j} \frac{\cos \theta_{ijk}}{2} \right) + \nonumber \\
&&\hspace{5.5cm} +\mathcal{O}(r_{ij}) \ . \nonumber
\end{eqnarray}
Using the expression above we can evaluate the matrix element of the commutator expression \eqref{FHcommut}, where the delta function restricts the spatial integration to short inter-particle separations
\begin{widetext}
\begin{align}
 \left \langle \chi \left|  g\sum_{i<j}  \delta \left( {\bf r}_{ij} \right) \frac{\partial \tau }{\partial r_{ij}} r_{ij} \right| \phi \right \rangle =& g\sum_{i<j}  {\Bigg [} - \left \langle \chi \left|   \delta \left( {\bf r}_{ij} \right)  \right| \phi \right \rangle  
-\sum_l^{i<l \ne j}  \left \langle \chi \left|   \frac{\delta \left( {\bf r}_{ij} \right) r_{ij}\cos \theta_{jil}}{2 r_{il}}  \right| \phi \right \rangle - \label{potmatel} \\
&\hspace{1cm} -\sum_l^{i \ne k < j}  \left \langle \chi \left|   \frac{\delta \left( {\bf r}_{ij} \right) r_{ij}\cos \theta_{ijk}}{2 r_{kj}}  \right| \phi \right \rangle  +\left \langle \chi \left|   \delta \left( {\bf r}_{ij} \right)  r_{ij} \mathcal{O}(r_{ij}) \right| \phi \right \rangle - \nonumber \\
& \hspace{2cm} -  \left( \frac{1}{a_s}+\sum_l^{i<l \ne j} \frac{\cos \theta_{jil}}{2a_s}  +\sum_k^{i \ne k < j} \frac{\cos \theta_{ijk}}{2a_s} \right) \left \langle \chi \left|   \delta \left( {\bf r}_{ij} \right)  r_{ij} \right| \phi \right \rangle {\Bigg ]} \ .
\nonumber
\end{align}
\end{widetext}
Assuming that the functions $\chi$ and $\phi$ are bounded,
the last two terms in \rref{potmatel} are zero as
 $\delta \left( {\bf r}_{ij} \right)  r_{ij}$
gives zero after performing the integral either for ${\bf r}_i$ or ${\bf r}_j$. Although the integrands in the second and the 
third term on the right-hand-side of \rref{potmatel} can be
finite at the coalescence points ${\bf r}_i={\bf r}_l$ and ${\bf r}_k={\bf r}_j$, they are still zero everywhere else. Since the coalescence points form a set of measure zero, these terms yield zero after integrating over the remaining variables. This leads to a matrix element of the Dirac delta function:
\begin{align}
& \left \langle \chi \left|  g\sum_{i<j}  \delta \left( {\bf r}_{ij} \right)  \frac{\partial \tau }{\partial r_{ij}} r_{ij} \right| \phi \right \rangle = \label{commmatrixelm}\\
&\hspace{4cm} =-g\sum_{i<j} \left \langle \chi \left|   \delta \left( {\bf r}_{ij} \right)  \right| \phi \right \rangle . \nonumber
\end{align}
Due to the bounded nature of the functions $\phi$ and $\chi$, 
the matrix element of the Fermi-Huang pseudopotential also reduces to the matrix element of the Dirac-delta function, but with the opposite sign, 
\begin{align}
& \left \langle \chi \left|  g\sum_{i<j}  \delta \left( {\bf r}_{ij} \right) \frac{\partial}{\partial r_{ij}} r_{ij} \right| \phi \right \rangle = \label{FHmatrixelm} \\
& \hspace{2cm}= g\sum_{i<j} \left \langle \chi \left|   \delta \left( {\bf r}_{ij} \right)  \right| \phi \right \rangle + \nonumber \\
 & \hspace{3cm} +g\sum_{i<j} \underbrace{\left \langle \chi \left|  \delta \left( {\bf r}_{ij} \right) r_{ij} \frac{\partial}{\partial r_{ij}} \right| \phi \right \rangle}_0 \ , \nonumber
\end{align}
where we have assumed that $\chi$ and $\partial \phi / \partial r_{ij}$ are bounded. Equation (\ref{FHmatrixelm}) shows that a matrix representation of the  the (physically meaningful) Fermi-Huang pseudopotential with sufficiently smooth (and bounded) basis functions is equivalent to the bare Dirac-delta pseudopotential, which is pathological in the sense that the infinite basis set limit does not exist.
After the transcorrelated similarity transformation, however, we obtain the two matrix elements \rrefsb{commmatrixelm} and \rrefsb{FHmatrixelm}, which cancel each other and thus eliminate
the irregular behavior in the matrix representation. Combining Eqs.\ \eqref{eq:trFH}, \eqref{FHcommut}, \eqref{commmatrixelm} and \eqref{FHmatrixelm} we finally obtain
\begin{align}
    \left \langle \chi \left| e^{-\tau} V_{\mathrm{FH}} e^{\tau}  \right| \phi \right \rangle&=    \left \langle \chi \left| V_{\mathrm{FH}} +\left[  V_{\mathrm{FH}} , \tau \right]  \right| \phi \right \rangle = 0\ .
\end{align}

\section{Smoothness of the transcorrelated eigenfunction for two particles \label{sec:appendixC}}

In this section we investigate the transcorrelated eigenfunction for two bosons or distinguishable particles with the same mass (e.g.\ fermions with different spin quantum number). We show that the singularity is reduced in the transcorrelated Hamiltonian due to the similarity transformation. Consequently, the transcorrelated eigenfunctions are not singular, there is only a cusp at the particle-particle coalescence point.

We consider the two-particle Hamiltonian  without trapping potential ($V_{\mathrm{trap}}=0$),
\begin{eqnarray}
H =& -\frac{\hbar^2}{2m} \nabla_\uparrow^2  
-\frac{\hbar^2}{2m} \nabla_\downarrow^2 + g \delta \left( {\bf r}_\uparrow - {\bf r}_\downarrow  \right) 
\frac{\partial}{\partial \left| {\bf r}_\uparrow - {\bf r}_\downarrow  \right| } 
\left| {\bf r}_\uparrow - {\bf r}_\downarrow  \right| \ , \nonumber 
\end{eqnarray}
where $\uparrow$ and $\downarrow$ label the two particles. 
Separating the center-of-mass from the relative motion coordinates, we obtain,
\begin{eqnarray}
H_\mathrm{rel}= -\frac{\hbar^2}{2 \mu } \nabla^2 + g \delta \left( {\bf r}  \right) \frac{\partial}{\partial r}  r \ ,
\label{Hrel}
\end{eqnarray}
where ${\bf r}={\bf r}_\uparrow -{\bf r}_\downarrow$, $\mu=m/2$ and the center-of-mass is described by  free-particle motion. 

Applying the transcorrelated  similarity transformation to the relative-motion Hamiltoninan of Eq.\ \rrefsb{Hrel} yields 
\begin{align}
 \tilde{H}_\mathrm{rel}=& e^{-\tau} {H}_\mathrm{rel} e^{\tau} \ , \nonumber \\
 \tilde{H}_{\mathrm{rel}}=& -\frac{\hbar^2}{2 \mu } \nabla^2 - \frac{\hbar^2}{ \mu } \left[ \frac{1}{2}\nabla^2 \tau + \left(\nabla \tau \right) \nabla + \frac{1}{2} \left(\nabla \tau \right)^2 \right] + \nonumber \\
& \hspace{2cm} + g \delta \left( {\bf r}  \right) 
\left[ \frac{\partial}{\partial r }r+r \left( \frac{\partial \tau}{\partial r} \right) \right]\ .
\label{htreltau}
\end{align}
Using \rrefsa{tauexpofu} and \rrefsb{ushortrangeapp} $\tau$ can be given explicitly at small interparticle separation,
\begin{align}
\tau = 1 -\ln \left(\frac{r}{a_s} \right) -\frac{r}{a_s}+\mathcal{O}\left( r^2\right) \ ,
\end{align}
with which the derivatives of $\tau$ in \rref{htreltau} can be expressed as
\begin{eqnarray}
\frac{\partial \tau}{\partial r} &=& -\frac{1}{r}-\frac{1}{a_s} + \mathcal{O}( r)  \ ,
\label{tauderr} \\
\nabla\tau  &=& - \frac{\bf{r}}{r^2}-\frac{\bf{r}}{a_s r} + \mathcal{O}({\bf r} ) \ , \label{nablatau} \\
\nabla^2 \tau &=&   - \frac{1}{r^2}-\frac{2}{a_s r}  + \mathcal{O}\left(r^0 \right) \ .
\label{nabla2tau}
\end{eqnarray}
Substituting back  into Eq.\ \rrefsb{htreltau}, an explicit expression for the Hamiltionian can be obtained for short distances
\begin{align*}
\tilde{H}_\mathrm{rel}= -\frac{\hbar^2}{2 \mu } \nabla^2   &+\frac{\hbar^2}{ \mu } \left(\frac{1}{r} + \frac{1}{a_s} \right)   \frac{\bf r}{r} \nabla + \\
& + g \delta \left( {\bf r}  \right) 
\left( r\frac{\partial}{\partial r }- \frac{r}{a_s} + \mathcal{O}\left( r^2\right) \right)
+ \mathcal{O}\left( r^0 \right)\ .
\end{align*}

In order to obtain the transcorrelated eigenfunction, let us substitute the Hamiltonian into the Schr\"odinger equation
\begin{align*}
 -\frac{\hbar^2}{2 \mu } \nabla^2 \phi& ({\bf r}) + \frac{\hbar^2}{ \mu } \left(\frac{1}{r} + \frac{1}{a_s} \right)  \frac{\bf r}{r} \nabla
 \phi({\bf r}) + \\
& + g \delta \left( {\bf r}  \right) \left( r\frac{\partial}{\partial r }- \frac{r}{a_s} + \mathcal{O}\left( r^2\right) \right)\phi({\bf r})
 = E' \phi({\bf r}), 
\end{align*}
where $E'=E+\mathcal{O}\left( r^0 \right)$. Due to the spherical symmetry we can transform the differential equation into polar coordinates and consider only $s$-wave solutions
\begin{align}
&-\frac{\partial^2 \phi(r)}{\partial r^2} +  \frac{2}{a_s} \frac{\partial \phi(r)}{\partial r}+   \label{diffeqrad} \\ 
& \hspace{0.5cm} + \frac{g \mu}{2 \pi \hbar^2 r} \delta \left( r  \right) 
 \left[ \frac{\partial}{\partial r }+ \frac{1}{a_s} +  \mathcal{O}\left( r^2\right) \right]\phi(r) =  \frac{2\mu E'}{\hbar^2} \phi(r) \ . \nonumber
\end{align}
The differential equation can be solved for small interparticle separation,
\begin{eqnarray*}
\phi(r) \stackrel{r \rightarrow 0}{=} e^\frac{r}{a_s}\left(c_1 e^{r\sqrt{\frac{\hbar^2}{2\mu a_s^2}-E'}}+c_2 e^{-r\sqrt{\frac{\hbar^2}{2 \mu a_s^2}-E'}}\right) \ ,
\end{eqnarray*}
where $c_1$, $c_2$ and $E'$ can be determined only if we know the solution in the whole space. 

Differentiating the wave function we  notice that it has a linear term,
\begin{eqnarray*}
\phi(r)  &=& c_1+c_2+b r+\mathcal{O}(r^2) \ ,
\end{eqnarray*}
where the prefactor $b$, before the linear term is
\begin{eqnarray*}
b=\phi'(0) &=& \frac{c_1+c_2}{a_s}+(c_1-c_2)\sqrt{\frac{\hbar^2}{2 \mu a_s^2}-E'} \ .
\end{eqnarray*}
Considering the spherical symmetry, we obtain a function which goes linearly to $c_1+c_2$ around the origin and forms a cusp. This function is not-singular and continuous, however, its first derivative is discontinuous. Therefore, the transcorrelated transformation smoothes the wave function from a hyperbolic singularity ($\sim1/r$) to a cusp feature.

\section{Second quantized form of the transcorrelated Hamiltonian \label{sec:appendixD}}

In this section, we give an explicit expression for the second quantized form of the transcorrelated Hamiltonian in a rectangular box with periodic boundary conditions.

Starting from the full Hamiltonian of Eq.~(7) of the main text we write
\begin{align}
H = H_\mathrm{k} + \sum_i V_\mathrm{trap}(\mathbf{r}_i) + V_\mathrm{FH} ,
\end{align}
where 
\begin{align}
H_\mathrm{k} = \sum_i -\frac{\hbar^2}{2m} \nabla_i^2 ,
\end{align}
is the kinetic energy operator. Under the transcorrelated similarity transformation
\begin{align}
\nonumber
\tilde{H} & = e^{-\tau} H e^\tau =  e^{-\tau} H_\mathrm{k} e^\tau + \sum_i V_\mathrm{trap}(\mathbf{r}_i) \\
& = \tilde{H}_\mathrm{k}+ \sum_i V_\mathrm{trap}(\mathbf{r}_i) ,
\end{align}
where we assume that will only apply the transcorrelated Hamiltonian in the domain of bounded and almost-everywhere differentiable functions, under which conditions the Fermi-Huang pseudopotential disappears according to Sec.~\ref{sec:trcorrFH}. The trapping potential is unchanged by the similarity transformation because it is a diagonal operator in coordinate space.

The transcorrelated kinetic energy operator obtains additional terms, as already discussed [see Eq.\ \eqref{trcorrHam}  in the main text]:
\begin{align} \nonumber
& \tilde{H}_\mathrm{k}  = -\frac{\hbar^2}{2m} \sum_i \left[ \nabla_i^2  + \frac{1}{2} \nabla_i^2 \tau  + \left( \nabla_i \tau \right) \nabla_i + \frac{1}{2} \left( \nabla_i \tau \right)^2 \right] .
\end{align}
Assuming a box with side length $L$ and periodic boundary conditions, we can introduce the usual plane-wave single-particle basis functions. Following the description in Ref.\ \cite{luo_combining_2018}, the second quantized form of the transcorrelated kinetic energy operator is easily determined as
\begin{widetext}
\begin{eqnarray}
\tilde{H}_\mathrm{k} &=& \frac{\hbar^2}{2m}\sum_{ {\bf k} \sigma }k^2 \, a_{{\bf k},\sigma}^{\dagger} \, a_{{\bf k},\sigma}  +  \sum_{\substack{{\bf pqk} \\ \sigma \sigma'}
} \, T_{\bf pqk} \Theta_{\sigma \sigma'} \,
a_{{\bf p}-{\bf k},\sigma}^{\dagger} \,  a_{{\bf q}+{\bf k},\sigma'}^{\dagger} \, a_{{\bf q},\sigma'} \, a_{{\bf p},\sigma}  \label{secondeffHam} \\ 
&& \hspace{7cm} +  \sum_{\substack{{\bf pqs} \\ {\bf kk'} \\ \sigma \sigma'}} 
Q_{\bf kk'}   \Theta_{\sigma \sigma'}  a_{\bp-\bk ,\sigma}^{\dagger} a_{\bq+\bkp,\sigma}^{\dagger} 
a_{\bs+\bk-\bkp,\sigma'}^{\dagger}
a_{\bs,\sigma'} a_{\bq, \sigma} a_{\bp, \sigma} \ , \nonumber
\end{eqnarray}
\end{widetext}
where $a_{\bk, \sigma}^\dagger$ creates a one-particle plane wave state with
momentum $\bk$ and spin $\sigma$
and $\Theta_{\sigma \sigma'}=\delta_{\sigma \sigma'}$ for bosons and
$\Theta_{\sigma \sigma'}=1-\delta_{\sigma \sigma'}$ for fermions. The tensors $\mathbf{T}$ and $\mathbf{Q}$ can be expressed explicitly as
\begin{eqnarray}
T_{\bf pqk} &=& 
 \frac{\hbar^2}{mL^3} \left( k^2 \tilde{u}(k) - (\bp-\bq)\bk\tilde{u}(k)+ \frac{W(\bk)}{L^3}  \right) \ , \nonumber \\
W(\bk) &=& \sum_{\bkp}(\bk-\bkp)\bkp\tilde{u}(\left| \bk-\bkp \right|)\tilde{u}(k') \ , \label{tensorT} \\
Q_{\bf kk'} &=& -\frac{\bkp \bk \tilde{u}(k) \tilde{u}(k')\hbar^2}{2mL^6} \ . \nonumber
\end{eqnarray}

\section{Numerical evaluation of the infinite summation in \rref{tensorT}} \label{sec:appendixE}

In this section we describe the algorithm that we have used for evaluating the infinite sum in \rref{tensorT}. 

First, let us realize that we can restrict the indices in the summation from below due to momentum cutoff in the correlation factor \rrefsb{corrfactapprox},
\begin{eqnarray}
W(\bk) &=& \sum_{\bkp}^{k',|\bk-\bkp| \ge k_c }(\bk-\bkp)\bkp\tilde{u}(\left|\bk-\bkp \right|)\tilde{u}(k') \ . \label{Wsum}
\end{eqnarray}
As we can see in \rref{Wsum}, the summation goes to infinity, which prohibits the exact evaluation. However, an accurate approximate value can be obtained if we partition the summation in \rref{Wsum} to a summation inside a sphere with radius $\kint$ and a summation outside this sphere,
\begin{eqnarray}
W(\bk) &=& P(\bk,\kint) + R(\bk,\kint) , \label{Wexp} \\
P(\bk,\kint) &=& \sum_{\bkp}^{\kint > k',|\bk-\bkp| \ge k_c } w(\bk,\bkp) \ , \\
R(\bk,\kint) &=& \sum_{\bkp}^{ k',|\bk-\bkp| \ge \kint } w(\bk,\bkp)\ , \\
w(\bk,\bkp) &=& (\bk-\bkp)\bkp\tilde{u}(\left|\bk-\bkp\right|)\tilde{u}(k') \ . \label{wexp}
\end{eqnarray}
As $w(\bk,\bkp)$ decays with $k^{-6}$ at larger values of $\bk$, we can approximate the summation with an integral in $ R(\bk,\kint)$,
\begin{align}
 R(\bk,\kint)  & \approx  R_I(\bk,\kint) \ , \nonumber\\
 R_I(\bk,\kint)& = \int\limits_{\kint}^\infty k'^2 \mbox{d}k' \int\limits_{-1}^1 \mbox{d}(\cos\theta)  \frac{\left(k k' \cos\theta - k'^2 \right)}{4 \pi^2} \cdot \nonumber \\ 
	& \hspace{0.8cm} \cdot \tilde{u}(\sqrt{k^2 -2kk'\cos\theta + k'^2})\tilde{u}(k') \ . \label{eq:integralexp}
\end{align}
Due to the additional conditions in the sum \rrefsb{Wsum}, further restrictions apply at the boundaries of the integral, when
\begin{eqnarray}
\sqrt{k^2 -2kk'\cos\theta + k'^2} &<& k_c \ . \label{kintcondition} 
\end{eqnarray}

In order to avoid the complicated limits of the integration, we choose $\kint$ large enough such that \rrefsb{kintcondition} never occurs. After some algebra, it can be shown that it is sufficient to choose  $\kint$ such as to satisfy
\begin{eqnarray}
\kint &\ge &  k + k_c \ ,
\end{eqnarray}
which is easy enough to fulfil as $k$ and $k_c$  is kept small to limit the size of the Hilbert-space and to enhance the effect of the correlation factor.  

In order to evaluate the integral \rrefsb{eq:integralexp}, we consider the Taylor-expanded form of $\tilde{u}(k)$ in \rref{corrfactapprox}, 
\begin{align*}
R_I({\bf k},\kint)  =  -\frac{2 \pi^2}{k_{\mathrm{int}}} - \frac{8 \pi}{a_s k_\mathrm{int}^2} - & \frac{32}{3 a_s^2 k_\mathrm{int}^3}- \\  &-\frac{4k^2\pi}{3a_s k_\mathrm{int}^4}+ {\mathcal O} \left( a_s^{-2} k_\mathrm{int}^{-5}\right)\ .
\end{align*}

In this paper we specifically consider unitary interactions, where the integral \rrefsb{eq:integralexp} can be  evaluated exactly
\begin{align}
\lim_{a_s \rightarrow \pm \infty } R({\bf k},\kint)  \approx \lim_{a_s \rightarrow \pm \infty } R_I({\bf k},\kint)  =  -\frac{2 \pi^2}{k_{\mathrm{int}}} \ . \label{Riunit}
\end{align}

In this work we have applied \rrefsa{Wexp}-\rrefsb{wexp} and \rrefsb{Riunit} to evaluate the matrix element $W(\bk)$ for up to $k=16\pi/L$ with $\kint=1600 \pi/L$, where the finite sum $P(\bk,\kint)$ was evaluated exactly and the infinite sum $R(\bk,\kint)$ was replaced by the intergral $R_I(\bk,\kint)$.
Varying $k$ up to the maximal value of $16 \pi/L$ we found the uncertainties in the values of $W(\bk)$ only in the seventh and eighth significant digits. As the energy scales linearly  with the error in the matrix elements, the error should appear in the energy in the same order.
Moreover, the accuracy of the integral approximation was also checked numerically by comparing the energies from $\kint=1200 \pi/L$ and $\kint=1600 \pi/L$ calculations. We did not find any significant difference in examples of two, three and four fermions.

\begin{figure}
\begin{center}
\includegraphics[scale=0.3]{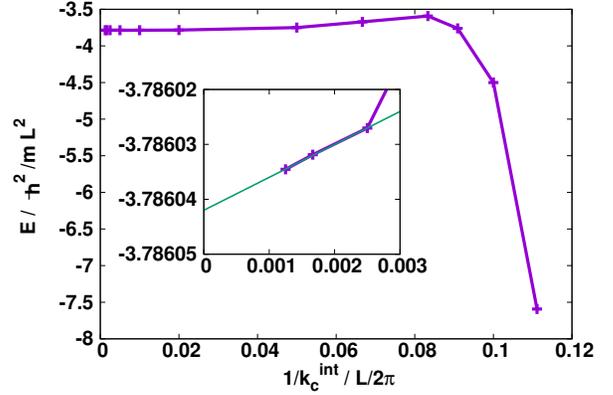}
\caption{Convergence of the energy of one spin-up and one spin-down particle at unitary interaction with $\kint$. The maximal values of the momentum for the single-particle basis was $16\pi/L$. The transcorrelated cutoff is kept to $k_c=2\pi/L$. The extrapolation to $1/\kint=0$ 
is determined with a linear fit to the last three data points. \label{fig:summation}}
\end{center}
\end{figure}

For two particles the convergence of the energy upon increasing $\kint$ is demonstrated in Fig.\ \ref{fig:summation}. The observed error seems adequate for our numerical calculations, where the uncertainty of our final results was in the fourth and fifth significant digits.

\section{Details of the numerical calculations \label{sec:appendixF}}

\subsection{FCIQMC}\label{sec:fciqmc}
For the numerical calculation we used the NECI code \cite{neci}, where  transcorrelated Hamiltonians including  three-body excitations had previously been  implemented for the homogeneous electron gas \cite{luo_combining_2018}, the Fermi-Hubbard model \cite{dobrautz_compact_2019}, atoms, molecules \cite{cohen_similarity_2019}, and the Fermi gas in one dimension \cite{jeszenszki_accelerating_2018}. In the context of this project we have further extended the capabilities of the NECI code by including the transcorrelated Hamiltonian for the unitary Fermi gas in three dimensions.

For two-particles, non-hermitian exact (deterministic) diagonalization is applied in NECI using an external Lapack library \cite{anderson_lapack_1999}. 
For three and four fermions the Hilbert-space is too large for deterministic diagonalization.  Hence the Full Configuration Interaction Quantum Monte Carlo (FCIQMC) algorithm \cite{booth_fermion_2009,booth_linear-scaling_2014} is applied to obtain the ground-state energy. 

One of the elementary parameters of the FCIQMC algorithm is the number of the walkers \cite{booth_fermion_2009,booth_linear-scaling_2014}. It controls the resolution of the wave function and the memory usage of the algorithm. In this algorithm a minimal number of walkers is required to eliminate the sign-problem.

\begin{figure}
\begin{center}
\includegraphics[scale=0.7]{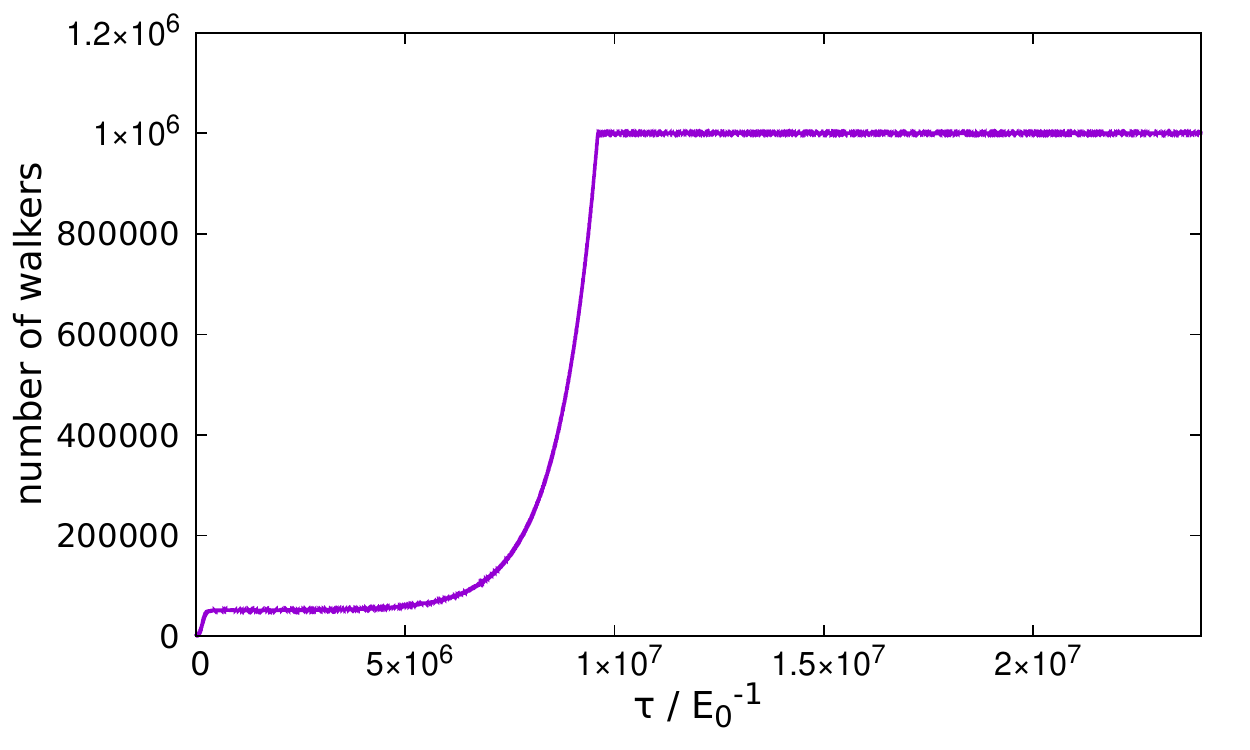}
\includegraphics[scale=0.7]{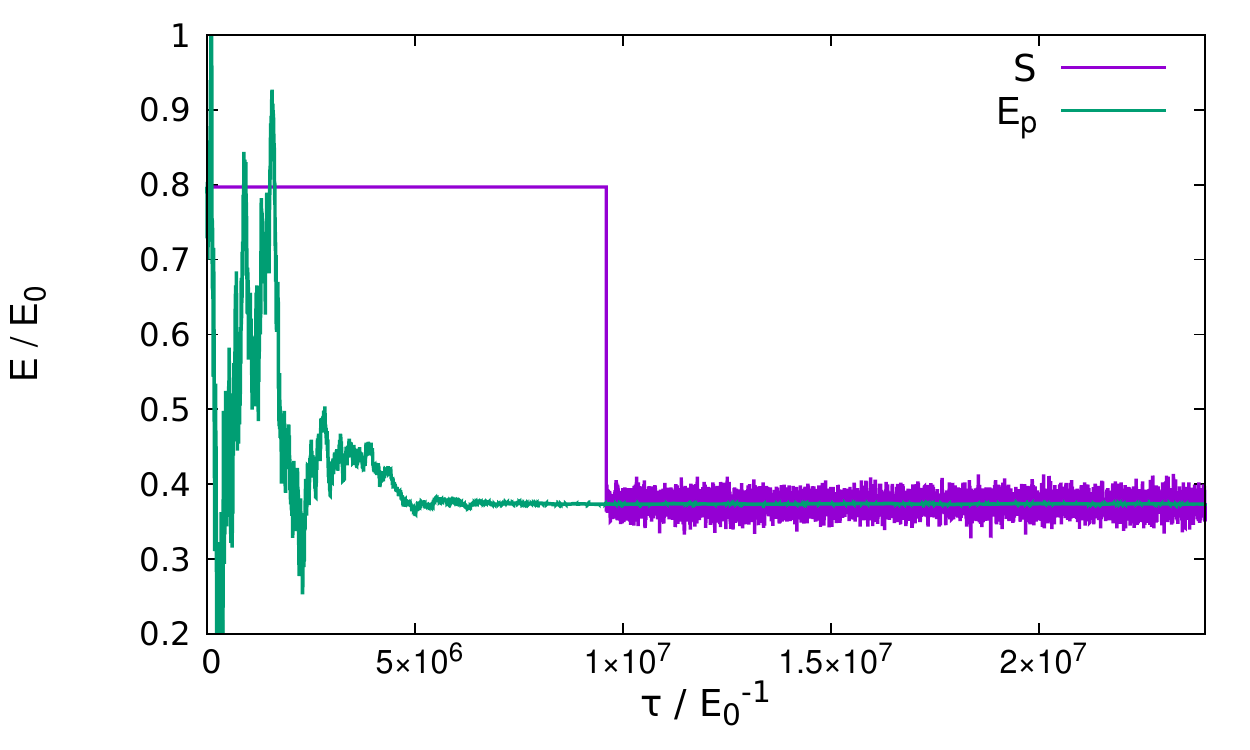}
\caption{The total number of walkers (top panel), and the shift $S$ and projected energy $E_p$ (bottom panel) during the FCIQMC simulation for the example of two spin-up and one spin-down particles with unitary interactions at $M=11^3$ using the transcorrelated approach at $k_c=2\pi/L$.  After the target walker number of $10^6$ is reached, the previously constant shift parameter is updated in order to control the walker number.
$E_0=4\pi^2\hbar^2/mL^2$ is the non-interacting energy. \label{fig:popdym}}
\end{center}
\end{figure}

The minimal number is determined by the annihilation plateau \cite{booth_fermion_2009, spencer_sign_2012}, which appears in the number of walkers during the imaginary time evolution. This plateau can be seen to appear in Fig.\ \ref{fig:popdym} at around 50,000 walkers. At the end of the plateau, around $\tau \approx 5\times 10^6 E_0^{-1}$, the sign structure of the wave function is determined, fluctuations in the projected energy $E_p$ are greatly reduced, and the walker number starts growing exponentially. When the number of the walkers exceeds the initially set target walker number of $10^6$, we start adjusting the initially constant shift parameter $S$ according to protocol of Ref.\ \cite{booth_fermion_2009} in order to control the walker number, which will subsequently fluctuate around a mean.
Both the shift parameter $S$  as well as the projected energy $E_p$ provide estimators for the ground state energy. The final value of the ground state energy is determined by the mean of shift parameter (after reaching the final walker number). The error is obtained from an estimate of standard deviation of the mean using a standard blocking analysis to remove auto-correlations in the time series \cite{Flyvbjerg1989}.

For all calculations for three fermions and for the lattice-renormalized calculations for four fermions , we were able to apply a large enough walker number to detect and exceed the annihilation plateau. However, for the transcorrelated calculations with four fermions with $M>9^3$ the annihilation plateau was too high for the available numerical resources. In these cases we applied the initiator method \cite{cleland_communications:_2010}, which has proved to be efficient for electronic structure calculations \cite{james_j._shepherd_investigation_2012,booth_towards_2012,li_manni_combining_2016}.  
While this approximation causes a systematic bias in the calculations, the bias disappears when  increasing the number of walkers.
For all results shown, the number of walkers was increased until the changes in energy were insignificant compared to the statistical error bars. Another systematic bias, the population control bias, is known to affect FCIQMC calculations with small walker number but is well below the statistical error for the parameters considered in our calculations. We thus expect the FCIQMC results presented in this work to be essentially free of any systematic bias.
 
 The parameters for the calculations are shown in Tables \ref{runparameters3p} and  \ref{runparameters4p}. The calculations were typically run on a single node with 20 or 40 processor cores for 3 to 9 days. The largest calculation was for four particles with $k_c=4\pi/L$ and $M=17^3$. The memory usage in this example was about 114 GByte and about 351 days of CPU time were used.  

\begin{table}
\begin{tabular}{|c|r|r|r|l|}
\hline
     $k_c / 2\pi L^{-1} $ & $M$ &  \multicolumn{1}{c|}{$N_w$} &  \multicolumn{1}{c|}{$N_\tau$} &  \multicolumn{1}{c|}{$\Delta \tau / E_0^{-1}$}  \\ \hline
     1 &    $11^3$  &  10000000 & 1310720 & 0.00010 \\
     1 &    $13^3$ &  16000000  & 327680 & 0.000067\\
     1 &    $15^3$ & 20000000 & 327680 &   0.000039\\
     2 &    $13^3$ & 2000000  & 1310720 & 0.00039 \\
     2 &    $15^3$ & 10000000 & 163840&   0.00018\\
     2 &    $17^3$ & 20000000 & 327680 &  0.00011 \\ \hline
     \end{tabular}
     \caption{Parameters of the FCIQMC calculation for two spin-up and one spin-down particles. $N_w$ is the number of walkers and $N_\tau$ is the number of time steps. The time step size $\Delta \tau$ was determined by the histogram tau search algorithm \cite{neci}. 
     An annihilation plateau was detected for all three-particle calculations.
 \label{runparameters3p}}
\end{table}

\begin{table}
\begin{tabular}{|c|r|r|r|l|}
\hline
     $k_c / 2\pi L^{-1}$ & $M$ &  \multicolumn{1}{c|}{$N_w$} &  \multicolumn{1}{c|}{$N_\tau$} &  \multicolumn{1}{c|}{$\Delta \tau  / E_0^{-1}$}  \\ \hline
     1 &    $9^3$  &  2000000 & 655360 & 0.000079 \\
     1 &    $11^3$ &  8000000  & 1310720 & 0.000051\\
     1 &    $13^3$ & 16000000 & 1310720 & 0.0000079\\
     2 &    $13^3$ & 9500000  & 1310720 & 0.0000079 \\
     2 &    $15^3$ & 16000000 & 327680&  0.000051\\
     2 &    $17^3$ & 160000000 & 327680 &  0.000029 \\ \hline
     \end{tabular}
     \caption{Parameters of the FCIQMC calculation for two spin-up and two spin-down particles, as in Table \ref{runparameters3p}. 
     An annihilation plateau was detected for the calculation with $M= 9^3$. All other calculations were performed using the initiator approach \cite{cleland_communications:_2010}.
     \label{runparameters4p}}
\end{table}

\subsection{Complete basis limit and uncertainty} \label{sec:extrapol}

\begin{figure}
\begin{center}
\includegraphics[scale=0.7]{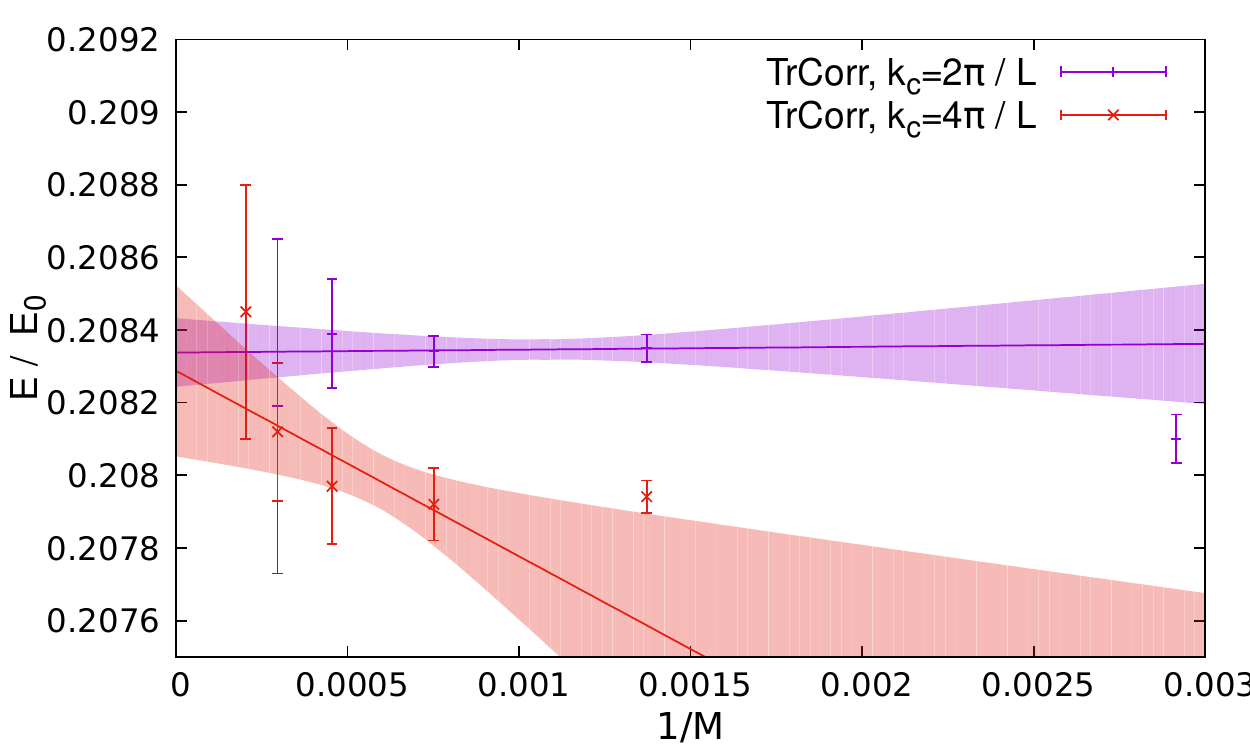}
\caption{The ground-state energy of two spin-up and two spin-down fermions. The purple and red bands show the $1\sigma$ confidence band obtained from $\chi^2$ fitting ($M= 9^3, 11^3,13^3, 15^3$ for $k_c=2\pi/L$, and $M=11^3, 13^3, 15^3, 17^3$ for $k_c=4\pi/L$). $E_0=4\pi^2\hbar^2/mL^2$ is the lowest non-interacting energy in the zero momentum sector. \label{fig:4bapp}}
\end{center}
\end{figure}

The complete basis limit of the energy, $E_\mathrm{cb}$, can be determined by extrapolation assuming that the asymptotic scaling of the energy $E$ with the size of the basis set is known. According to the observed scaling of the two-particle energy seen in Fig.~\ref{fig:absenerg2part} and discussed in Sec.~\ref{sec:results}, we expect the basis set error of the energy to be inversely proportional to the number $M$ of plane wave modes. Thus
\begin{align}
E/E_0 &= \alpha  + \beta \frac{1}{M} \ , \label{linearparam}
\end{align}
where  $\alpha =  E_\mathrm{cb}/E_0$ and $\beta$ are dimensionless fitting parameters.
Linear fits to Eq.~\eqref{linearparam} are shown in Fig.~\ref{fig:3b} for three fermion data and in Fig.~\ref{fig:4bapp} for four-fermion data, where $\alpha$ and $\beta$ represent the slope and intersect of the fitted lines, respectively. 
Estimators for the mean values of $\alpha$ and $\beta$, their variances $\sigma_\alpha^2$ and $\sigma_\beta^2$, and their covariance $\mathrm{cov}(\alpha, \beta)$ are obtained  using chi-square fitting \cite{press_numerical_2007}. This assumes that each data point is a Gaussian random variable with standard deviation given by the error bar, as justified in Monte Carlo simulations. 

Since we assume the linear relationship \eqref{linearparam} to hold only asymptotically for large $M$, we have to decide which data points to include in the linear fit. We expect our calculations with the smaller cut-off parameter $k_c = 2\pi/L$ to enter the asymptotic regime for smaller $M$ compared to the larger value $k_c = 4\pi/L$ because of the larger correlation factor. We also expect simulations with both values of $k_c$ to share the same limit $E_\mathrm{cb}$, and thus independently extrapolate to the same intersect.  Thus we first consider the data for the smaller cut-off value  $k_c = 2\pi/L$ and choose the smallest value of $M$ above which all data points for the energy $E$ reasonably represent the linear relationship \eqref{linearparam}. For Fig.~\ref{fig:3b} this corresponds to $M=9^3$ ($1/M\approx 0.0014$) and yields four data points. Then we consider the data for the larger value $k_c = 4\pi/L$ and choose the largest $M$ such that the intersects for interpolation with both cutoff values are consistent within the one-$\sigma$ confidence interval. For Fig.~\ref{fig:3b} this yields three data points at  $k_c = 4\pi/L$ with $M\ge 11^3$. 
For the four-fermion data in Fig.~\ref{fig:4bapp} the same procedure yields four data points for $k_c = 2\pi/L$ and four data points for $k_c = 4\pi/L$. 
It can be seen from the figures that the procedure is successful and  the intercepts have overlapping confidence intervals.
For both three- and four-fermion data sets, the extrapolations with the  smaller cut-off value  $k_c = 2\pi/L$ yield the smaller confidence intervals for the complete basis set limit $E_\mathrm{cb}$, and thus the corresponding values are reported as the final results.

The results of the complete basis set extrapolation for four fermion calculations are compared with literature results in Fig.\  \ref{fig:4b}  and  in Table \ref{tab:fourparticle}.

The error bands shown in Figs.~\ref{fig:3b} and \ref{fig:4bapp} were calculated as the $1\sigma$ confidence intervals for the energy  for each value of $1/M$ using the following procedure.
The $1\sigma$ confidence interval at the given value of $x=1/M$ can be calculated from
the standard deviation $\sigma_a(x)$ of the intercept $a$ in the linear equation
\begin{align}
  E/E_0 = a + b\left( \frac{1}{M} - x \right) .   
\end{align}
The $68\%$ or $1\sigma$ confidence interval is then given as 
\begin{align}
a - \sigma_a(x) <\frac{E}{E_0} < a + \sigma_a(x).
\end{align}
The new parameters $a$ and $b$ can be expressed by the original parameters in \rref{linearparam},
\begin{align}
a&=\alpha + \beta x , \label{bexpression} \\
b&=\beta . 
\end{align}
Using \rref{bexpression}, the standard deviation of parameter $a$ is then determined as \cite{press_numerical_2007}
\begin{align}
 \sigma_a(x)&= \sqrt{ \sigma_{\alpha}^2 + 2\, \mathrm{cov}(\alpha, \beta) x +  \sigma_{\beta}^2 x^2}. 
\end{align}

\begin{table}[]
    \centering
    \begin{tabular}{|c|l|l|}
    \hline
          Method       &  $E / E_0$ & $\mathrm{SE}(E/E_0)$\\ \hline
        Transcorrelated FCIQMC & $0.208338$ & $0.000094$ \\
        Hubbard FCIQMC & $0.2087$ & $0.0021$\\
         Quadratic dispersion FCIQMC & $0.2087$ &  $0.0011$ \\
          Endres 1 AFQMC $\mathcal{O}(4)$ \cite{endres_lattice_2013}  & 0.2122 & 0.0040 \\
         Endres 2 AFQMC $\mathcal{O}(5)$ \cite{endres_lattice_2013}  & 0.2130 & 0.0026 \\ 
         Bour 1 AFQMC \cite{bour_precision_2011} & 0.211 & 0.002 \\
         Bour 2 AFQMC 2 \cite{bour_precision_2011} & 0.210 & 0.002 \\
         Bour 3 AFQMC Euclidian \cite{bour_precision_2011} & 0.206 & 0.009 \\ 
         Yin ECG \cite{yin_small_2013} & 0.2058 & 0.0021 \\ \hline
    \end{tabular}
    \caption{Numerical values of the data shown in Fig.\  \ref{fig:4b}. Ground-state energies for two spin-up and two spin-down particles. The renormalization methodology for the different dispersion relations is described in Appendix \ref{sec:disp} and follows Refs.\ \cite{werner_general_2012,Carlson2011}.}     \label{tab:fourparticle}
\end{table}

\subsection{Extrapolated values for 4 particles in renormalized lattice calculations} \label{sec:4val}
\begin{figure}
\begin{center}
\includegraphics[scale=0.6]{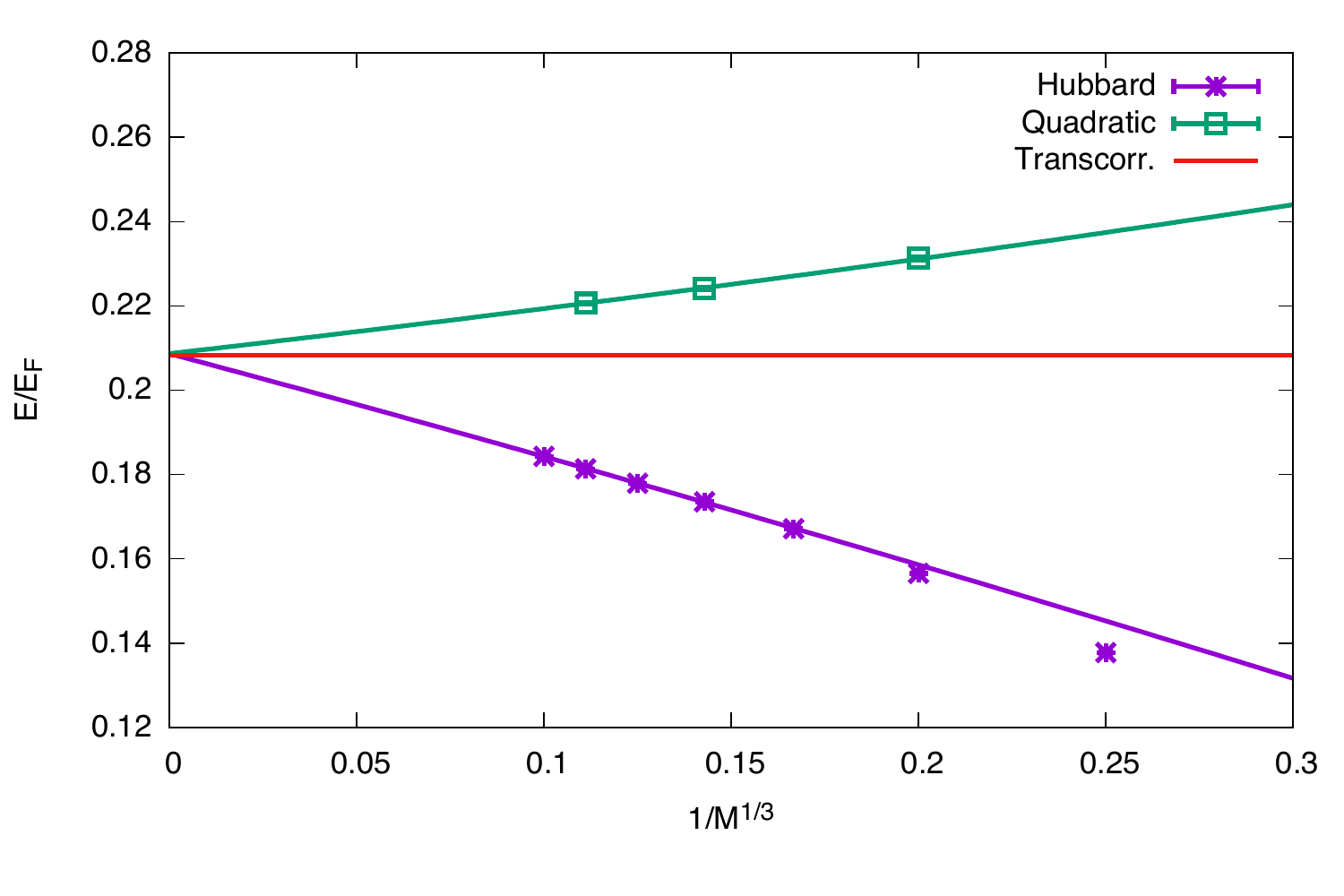}
\caption{Fitting procedure for results obtained using the renormalized lattice method with different single-particle dispersions. The value obtained using the transcorrelated method is included for comparison as the red line, with the error being smaller than the line width is this plot. The quadratic dispersion leads to two sets of points, depending on whether $M$ is even or odd. Here, we only show the results for odd values, which are considerably closer to the final extrapolated result.}
\label{fig:4blattice}
\end{center}
\end{figure}

In the main text we present extrapolated values for the ground state energy for four fermions obtained using the renormalized lattice method with the standard Hubbard dispersion and a quadratic dispersion in Fig.\  \ref{fig:4b}. These dispersions lead to a dominant convergence rate proportional to $M^{-1/3}$, as can be seen in Fig.\ \ref{fig:4blattice}. We fit a function $f(M)=\frac{E}{E_0}+A M^{-1/3}+BM^{-2/3}$ into the FCIQMC results to obtain an extrapolated value for $E/E_0$ with fitting error.

\subsection{Dispersions for renormalized lattice calculations}\label{sec:disp}
A useful basis set expansion for a free space Fermi system is using the Hubbard model in the low density limit. In this regime, any single-particle dispersion which is quadratic around zero will converge to the same result in the infinite-basis states (lattice sites) limit \cite{pricoupenko_three_2007}. Werner and Castin have proposed to replace the standard Hubbard single-particle dispersion with several candidates which eliminate finite effective-range contributions to varying degrees \cite{werner_general_2012} This leads to a convergence rate improvement from $M^{-1/3}$ for the Hubbard and quadratic dispersions to $M^{-2/3}$ for so-called "magic" dispersions. While a quartic dispersion was fitted to converge with the same power law as the magic dispersion in ref.~\cite{Carlson2011}, for the smaller particle numbers in our case we find a small contribution remaining that is proportional to $M^{-1/3}$. In Table \ref{tab:dispersions}, we list all dispersions used in this work with the renormalized Hubbard interaction parameters corresponding to unitary interactions.

\begin{table}[]
    \centering
    \begin{tabular}{|c|l|l|}
    \hline
        Name & Function & $U/t$ \\
    \hline
        Hubbard & $2t\sum_{i=1}^3(1-\cos(k_i\alpha))$ & -7.91355 \\
        Quadratic & $t(\vec k\alpha)^2$ & -10.28871\\
        Quartic & $t(\vec k\alpha)^2(1-C_1 \left(\tfrac{\vec k \alpha}{\pi}\right)^2]$ & -8.66661 \\
        Magic & $12tX(1+C_2X+C_3X^2)$ & -12.89076 \\
    \hline
    \end{tabular}
    \caption{Single-particle dispersions as functions of lattice momentum $\vec k$ used in FCIQMC-simulations in this work for comparison with the transcorrelated method. Here, $U$ and $t$ denote the usual Hubbard interaction and hopping parameters, $\alpha$ is the lattice constant. The numerical constants are $C_1=0.257022$, $C_2=-12.89076$, $C_3=-1.728219$, and $X=\frac16(\sum_{i=1}^31-\cos(k_\alpha))$. All functions and numerical values can be found with more details in Ref.~\cite{werner_general_2012}.}
    \label{tab:dispersions}
\end{table}

\subsection{Effective three-body interaction terms}\label{sec:threebodyterms}

Evaluating the transcorrelated Hamiltonian during any diagonalization procedure requires increased numerical effort compared to the renormalized lattice Hamiltonian. The largest part of the increased effort can be attributed to the three-body term and thus scales with $N^3$, where $N$ is the number of particles. An efficient procedure for implementing the three-body term in the FCIQMC algorithm is described in Ref.~\cite{dobrautz_compact_2019}. By sampling the three-body interactions at a lower rate than the much stronger and more important two-body interactions, the increased computational effort for having more non-zero off-diagonal matrix elements can be mostly mitigated, leaving a modest additional cost for treating the three-body-interaction terms explicitly. 

The results presented in the main part of the paper were computed by fully including all three-body excitations. In the following we discuss an approximate procedure previously used in Refs.~\cite{jeszenszki_accelerating_2018,luo_combining_2018} that only requires evaluating effective two-body matrix elements, and thus reduces the numerical effort further, while still producing highly accurate results. Specifically, for our three-particle calculations the computation time reduced by a factor of 2 -- 3  with approximated three body terms (effective two-body interactions only) compared to explicitly evaluating all three-body terms. For the four-particle calculation the speed-up factor was approximately 3 -- 4.

The approximate three-body interactions work by only allowing  excitations that change not more than two single-particle orbitals in the Fock-state $| \Phi \rangle$
\begin{widetext}
\begin{eqnarray} \label{eq:approx3b}
\sum_{\substack{{\bf pqs} \\ {\bf kk'} \\ \sigma}} 
Q_{\bf kk'}   a_{\bp-\bk ,\sigma}^{\dagger} a_{\bq+\bkp,\sigma}^{\dagger} 
a_{\bs+\bk-\bkp,\bar{\sigma}}^{\dagger}
a_{\bs,\bar{\sigma}} a_{\bq, \sigma} a_{\bp, \sigma} | \Phi \rangle &\approx& 
\sum_{\substack{{\bf pq} \\ {\bf k} \sigma }} 
N_{\bar{\sigma}} Q_{\bf kk}   a_{\bp-\bk ,\sigma}^{\dagger} a_{\bq+\bk,\sigma}^{\dagger}  a_{\bq, \sigma} a_{\bp, \sigma} | \Phi \rangle - \\
\nonumber
&& \hspace{0.5cm}
-\sum_{\substack{{\bf ps} \\ {\bf k} \sigma }}   N_{\sigma} Q_{\bf p-q , k}   a_{\bp-\bk ,\sigma}^{\dagger} \left( a_{\bs+\bp-\bq+\bk,\bar{\sigma}}^{\dagger}  + a_{\bs+\bq-\bp+\bk,\bar{\sigma}}^{\dagger}\right)a_{\bs, \bar{\sigma}} a_{\bp, \sigma}  | \Phi \rangle  \ , 
\end{eqnarray}
 \end{widetext}
 where $\sigma \ne \bar{\sigma}$, $N_{\sigma}$ is the number of the particles with spin $\sigma$ and we used the identity $\sum_r a_{r,\sigma}^\dagger a_{r,\sigma} | \Phi \rangle = N_\sigma | \Phi \rangle$.
This approximation is closely related to the Random Phase Approximation (RPA) \cite{gaskell_collective_1961, armour_calculation_1980,fetter2003quantum}. 

\begin{figure}
\begin{center}
\includegraphics[scale=0.7]{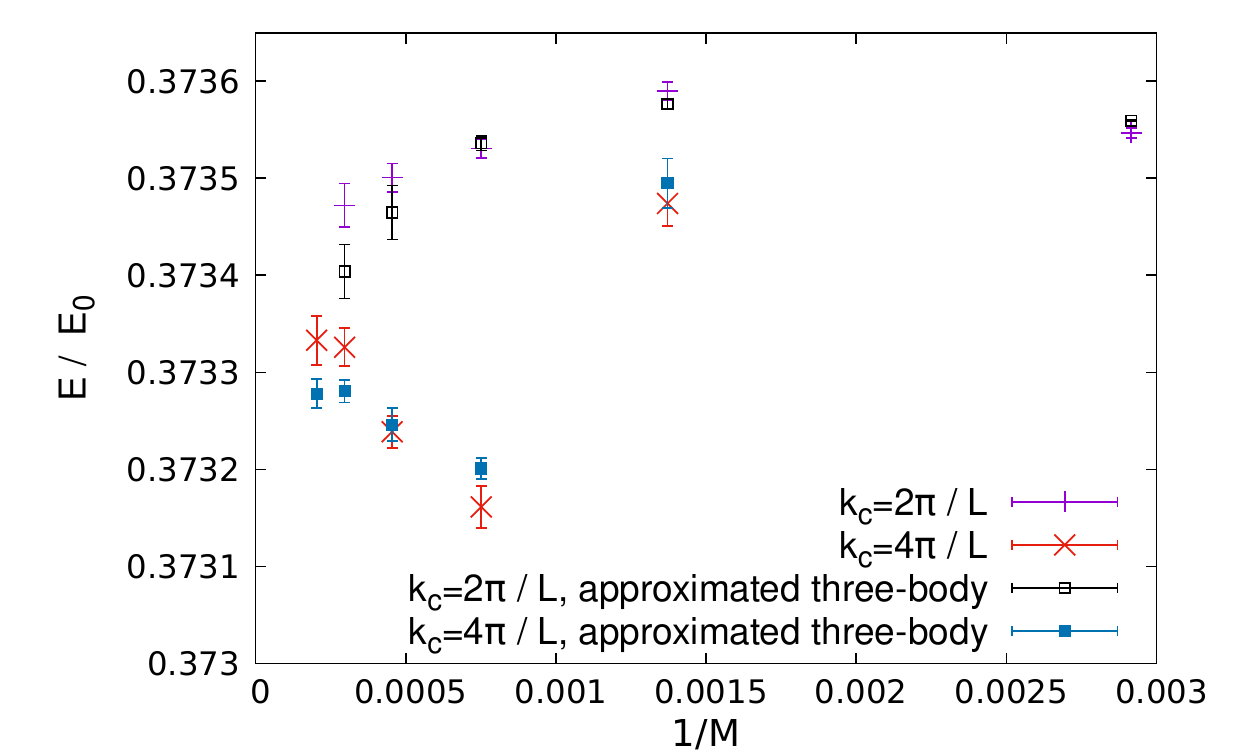}
\caption{The lowest energy of two spin-up and one spin-down fermions in the zero momentum sector with the full transcorrelated Hamiltonian and with the approximated three-body term as per Eq.\ \eqref{eq:approx3b}.
$E_0=4\pi^2\hbar^2/mL^2$ is the energy with zero interaction between the fermions. \label{fig:ttriples3b}}
\end{center}
\end{figure}

\begin{figure}
\begin{center}
\includegraphics[scale=0.7]{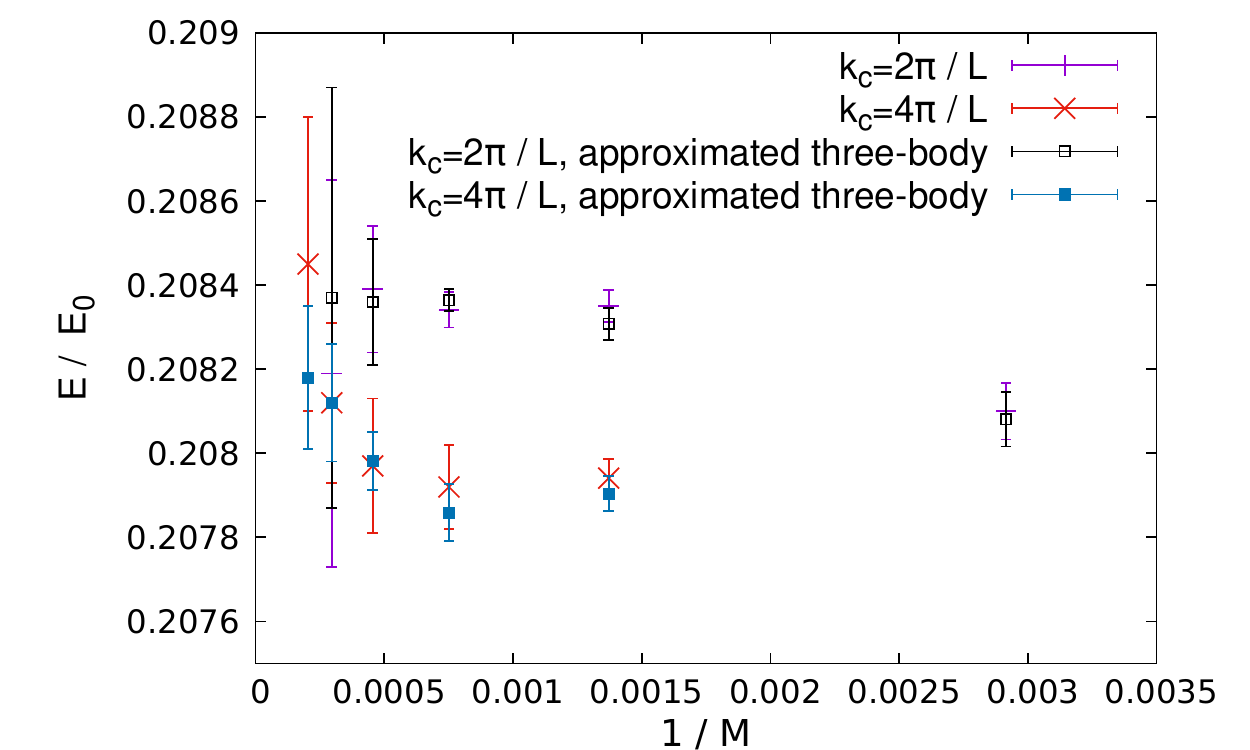}
\caption{The ground-state energy of two spin-up and two spin-down fermions with the full transcorrelated Hamiltonian and with the approximated three-body term  as per Eq.\ \eqref{eq:approx3b}.  $E_0=4\pi^2\hbar^2/mL^2$ is the noninteracting energy. \label{fig:triples4b}}
\end{center}
\end{figure}

Ground state energies computed with approximated three-body terms are compared to the full transcorrelated Hamiltonian in Fig.\ \ref{fig:ttriples3b} for three fermions and in Fig.\ \ref{fig:triples4b} for four fermions. We find that the approximate results and the full transcorrelated results have mostly overlapping Monte Carlo (statistical) error bars. The difference between full and approximated three-body terms is not statistically significant. We thus conclude that the error made by approximating the three-body terms with the RPA-like right hand side of Eq.~\eqref{eq:approx3b} is below the statistical Monte Carlo error for our calculations.

\bibliography{experiments,quantum_simulation,Bose_gas,book,Fermi_gases,Fewfermions,Fewbosons,FCIQMC,Bertsch_parameter,renormalization,QMC_ug,Bethe_Peiers,transcorrelated,scattering,math,programs,Fermi_Gas}

\end{document}